\documentclass[accepted=2026-27-02,a4paper,onecolumn,superscriptaddress]{quantumarticle}
\pdfoutput=1
\usepackage{graphicx} 
\usepackage{amsmath,mathtools,amssymb}
\usepackage{xparse}
\usepackage{physics}
\usepackage{microtype} 
\usepackage{tikz}
\usetikzlibrary{fit, positioning, calc}
\usepackage{amsfonts}
\usepackage{amsthm}
\usepackage{thmtools}
\usepackage{enumitem}
\usepackage{nicematrix}
\usepackage{booktabs,tabularx}
\usepackage{subcaption} 

\theoremstyle{definition} 

\newtheorem{definition}{Definition}

\newtheorem{technical}{Technical aside}

\newtheorem{notation}{Notational aside}
\newtheorem{multimode}{Multi-mode aside}

\usepackage{hyperref}
\usepackage[capitalize]{cleveref} 

\renewcommand{\H}{\mathcal{H}}

\newcommand{\dB}{{\mathbf{d_B}}}
\newcommand{\loss}{{\boldsymbol{\eta}}}
\newcommand{\cg}{\text{cg}} 
\renewcommand{\c}{{\text{c}}} 
\newcommand{\nc}{\text{nc}} 
\newcommand{\etamin}{\eta_{\min}}
\newcommand{\etamax}{\eta_{\max}}
\newcommand{\dmax}{d_{\max}}
\newcommand{\tmax}{t_{\max}} 
\newcommand{\tmin}{t_{\min}} 
\newcommand{\teff}{t^{*}} 
\newcommand{\teffmax}{\teff_{\max}} 
\newcommand{\teffmin}{\teff_{\min}} 

\newcommand{\idealEff}{\eta^*} 
\newcommand{\noise}{\text{noise}}
\newcommand{\ideal}{\text{ideal}}

\newcommand{\qbasis}{q_{>0}}

\newcommand{\GammancBasis}{\Gamma_{\nc,\; b}}
\newcommand{\GammazeroBasis}{\Gamma_{0,\; b}}
\newcommand{\GammaoneBasis}{\Gamma_{1,\; b}}

\newcommand{\epsSec}{\varepsilon_{\mathrm{sec}}} 
\newcommand{\epsAT}{\varepsilon_{\mathrm{AT}}} 
\newcommand{\epsPA}{\varepsilon_{\mathrm{PA}}} 

\newcommand{\epsIID}{\varepsilon_{\mathrm{IID}}}

\NewDocumentCommand\PP{gg}{
    \ensuremath{\mathcal{P}_{\IfNoValueTF{#2}{}{{#1 \vert #2}}}}}
\NewDocumentCommand\PPdb{gg}{
    \ensuremath{\mathcal{P}^{\dB}_{\IfNoValueTF{#2}{}{{#1 \vert #2}}}}}
\NewDocumentCommand\PPloss{gg}{
    \ensuremath{\mathcal{P}^{\loss}_{\IfNoValueTF{#2}{}{{#1 \vert #2}}}}}
\NewDocumentCommand\noiseChannel{g}{
    \ensuremath{\Phi_{\IfNoValueTF{#1}{\dB, \loss}{{#1}}}}}

\newcommand{\singleSet}{S}
\newcommand{\multiSet}{M}

\newcommand{\infPOVM}[1][]{\vec{\Gamma}_{#1}}
\newcommand{\finPOVM}[1][]{\vec{F}_{#1}}

\newcommand{\Msq}{{\PP}_{\text{sq}}}
\newcommand{\Mdc}{{\PP}_{\text{dc}}}
\newcommand{\Mcg}{{\PP}_{\cg}} 

\newcommand{\vac}{\text{vac}}

\newcommand{\nFSS}{N}
\newcommand{\nmeas}{{n_{\text{meas}}}}
\newcommand{\nS}{{n_S}} 
\newcommand{\nM}{{n_M}} 
\newcommand{\lambdamin}[1]{\lambda_{\min}\left(#1\right)} 

\newcommand{\classState}{\tau_{\text{C}}}

\newcommand{\rhodB}{\rho_{AQ}^\dB}

\newcommand{\subsubsubsection}[1]{%
  \par\vspace{0.75\baselineskip}%
  \begin{center}
    \bfseries #1%
  \end{center}
  \vspace{0.25\baselineskip}\noindent
}

\tikzset{
  pics/detector/.style args={#1,#2,#3}{
    code={
      \draw[line width=0.5mm] (0,-1) -- (0,1);
      \draw[line width=0.5mm] (0,1) -- ++(0.1,0);
      \draw[line width=0.5mm] (0,-1) -- ++(0.1,0);
      \draw[line width=0.5mm] (0.1,1) arc (90:-90:1); 
      \node[#3, align=center] (#1) at (0.5,0) {#2};
    }
  }
}

\tikzset{
  pics/source/.style args={#1,#2,#3,#4}{
    code={
    \coordinate (A) at (0,1);
    \coordinate (B) at (0,-1);
    \coordinate (C) at (3,-1);
    \coordinate (D) at (4,0);
    \coordinate (E) at (3,1);
    
    \draw[line width=0.4mm] (A) -- (B) -- (C) -- (D) -- (E) -- cycle; 
    \node[transform shape,scale=#4,#3, align=center] (#1) at (1.5,0) {#2};
    }
  }
}

\tikzstyle{process} = [rectangle, line width=0.3mm, minimum width=1cm, minimum height=1cm, text centered, draw=black]

\begin{document}

\title{Imperfect detectors for adversarial tasks with applications to quantum key distribution
}
\author{Shlok Nahar}
\affiliation{Institute for Quantum Computing and Department of Physics and Astronomy, University of Waterloo, Waterloo, Ontario, Canada, N2L 3G1}
\author{Devashish Tupkary}
\affiliation{Institute for Quantum Computing and Department of Physics and Astronomy, University of Waterloo, Waterloo, Ontario, Canada, N2L 3G1}
\author{Norbert L\"utkenhaus}
\affiliation{Institute for Quantum Computing and Department of Physics and Astronomy, University of Waterloo, Waterloo, Ontario, Canada, N2L 3G1}
\date{August 21, 2025}

\begin{abstract}
    Security analyses in quantum key distribution (QKD) and other adversarial quantum tasks often assume perfect device models. However, real-world implementations often deviate from these models. Thus, it is important to develop security proofs that account for such deviations from ideality.
    In this work, we extend the idea of squashing maps to develop a general framework for analysing imperfect threshold detectors, treating uncharacterised device parameters such as dark counts and detection efficiencies as adversarially controlled within some ranges. This approach enables a rigorous worst-case analysis with exactly characterised devices, ensuring security proofs remain valid under realistic conditions.
    Our results strengthen the connection between theoretical security and practical implementations by introducing a flexible framework for integrating detector imperfections into adversarial quantum protocols.
\end{abstract}

\maketitle

\section{Introduction}

Many analyses of quantum information (QI) processing tasks rely on exact models of the devices used, particularly in adversarial applications like quantum key distribution (QKD). However, as we near the practical implementation of QI protocols, it is of increasing importance to perform more refined analyses that take into account experimental imperfections. In particular, devices cannot be perfectly characterised, and thus the theoretical analysis must account for cases involving only partial models.
Moreover, for adversarial tasks such as QKD \cite{bennett1984quantum}, entanglement verification \cite{audenaert2006correlations}, and quantum-secure multiparty deep learning \cite{sulimany2024quantum}, the adversary might even have some limited control over the devices. This further exacerbates the need for more refined theoretical analyses in the adversarial setting.

This has led to growing attention to addressing imperfections in quantum devices, particularly in the context of QKD; both for realistic sources \cite{curras2023securityframework, arqand2024mutual, sixto2024quantum, zapatero2021security, curras2023security}, and for realistic detection setups \cite{fung2008security,lydersenSecurity2010,maroy2010security, tupkary2024phaseerrorrateestimation}.
While this body of work constitutes a tremendous amount of progress towards implementation security of QKD protocols, it is primarily restricted to the entropic uncertainty relation (EUR)- and phase-error correction-based proofs. On the other hand, work \cite{marwah2024proving, arqand2024mutual} addressing imperfections compatible with entropy accumulation theorem-based proofs either assume qubit sources \cite{marwah2024proving}, or require bounds on quantities that cannot be easily related to physical device parameters \cite{arqand2024mutual}. Thus, it is important to develop methods that are more broadly applicable; both across proof techniques within QKD as well as for applications outside QKD.

In this work we address the problem of imperfect detection setups with threshold detectors by extending the idea of squashing maps \cite{tsurumaru2010squash,beaudry2008squashing,gittsovich_squashing_2014,zhang_security_2021,tsurumaru2008security} in a general framework, leaving the problem of imperfect sources for future work.
Our framework, introduced in \cref{sec:proofTechnique}, treats imperfectly characterised device parameters equivalently to untrusted device parameters by `giving' the uncharacterised component to Eve. Thus, we reduce the analysis to a `worst-case' scenario with exactly characterised devices where existing techniques are applicable. Moreover, this approach naturally permits a controlled level of adversarial influence over the detection setup, thereby also proving security against bounded side-channel attacks.

We emphasise that formally accomplishing this task is non-trivial as noticed in \cite[Section V.B.]{li_improving_2020} for the case of dark counts, and thus a rigorous analysis is needed.
Furthermore, our framework relies solely on the intrinsic properties of the detection setup, without making any assumptions about the specific protocol in which the detection setup might be employed. Thus, we expect this to apply to adversarial tasks, in general.

We describe the techniques that allow for such a rigorous analysis for the case of dark counts and detector loss in \cref{subsec:noiseChannelDarkCounts,subsec:noiseChannelLoss} respectively.
Importantly, our results are presented in terms of experimentally measurable parameters, including the maximum dark count rate and the range of potential detector loss values.
We also extend our analysis in \cref{subsubsec:NoiseChannelGenSetups} to a form (\cref{thm:genericNoiseChannel}) that can be applied for arbitrary imperfections. Finally, we detail the application of our results to the various proof techniques for QKD in \cref{sec:appToQKD}. The impact of our improvements for QKD security proofs is summarised in \cref{tab:imperfect-threshold-setups}.

Memory effects are an important consideration in realistic detection setups, as all practical detectors exhibit phenomena such as dead times and afterpulsing. While most of our analysis assumes memoryless detectors, in \cref{sec:corrDetEAT} we sketch a proof idea for incorporating detector memory effects into a QKD security proof. Our approach is tailored to a specific proof technique — namely, the marginal-constrained entropy accumulation theorem (MEAT) developed in Ref.~\cite{arqand2025marginal} — and contributes an initial step toward the broader goal of incorporating detector memory effects into QKD security proofs.
We stress that our contribution is limited to a proof sketch, and that completing the full argument remains a technically challenging and important direction for future work. We also note that alternate ways to address this problem within phase error estimation-based proofs has been described in Ref.~\cite{wang2025phase}.
For clarity and accessibility, \cref{sec:corrDetEAT} is written to be self-contained and can be read independently of the rest of the paper.

\section{Detector Model} \label{sec:DetectorModel}

In this section, we describe a commonly used model 
for threshold detectors, and setups that utilise these detectors.
Threshold detectors are devices that ideally `click' in the presence of incident photons, and don't click in the absence of incident photons.
Deviations from ideality are termed dark counts, or detector efficiency as elaborated on in the following subsections.

\subsection{Dark counts as classical post-processing} \label{subsec:DCPP}

Dark counts are clicks that are registered by a detector in the absence of any incident photons. There are, broadly speaking, two mechanisms that explain this phenomenon.
Firstly, single-photon detectors sometimes have dark counts as a result of thermal noise, where thermally generated carriers or quantum tunnelling of electrons can cause spurious detections \cite{dao2024single}. This results in dark counts that are independent of any other click events. Secondly, after a click event, SPAD detectors sometimes have trapped charges that are released at a later time causing additional click events, commonly called afterpulsing \cite{migdall2013single}.
In this work, we primarily deal with the effect of dark counts that are independent of other click events, and each other. The effects of afterpulsing are considered only in \cref{sec:corrDetEAT}, where we defer the development of a detailed model.

The independent dark counts can be modelled as a classical post-processing being applied to the measurement outcomes. For example, consider a single threshold detector with dark count rate $d_B$ having POVM $\{\Gamma^{d_B}_{\nc}, \Gamma^{d_B}_{\c}\}$ corresponding to no-click and click events respectively.
This can be written in terms of the POVM $\{\Gamma_{\nc}, \Gamma_{\c}\}$ of a threshold detector without dark counts as
\begin{equation} \label{eq:DCPPforSingleDet}
    \begin{aligned}
        \Gamma^{d_B}_{\nc} &= (1-d_B)\Gamma_{\nc}\\
        \Gamma^{d_B}_{\c} &= \Gamma_{\c}+d_B\Gamma_{\nc}.
    \end{aligned}
\end{equation}

\Cref{eq:DCPPforSingleDet} can be more concisely written as 
\begin{align} \label{eq:DCPPforSingleDetVecVersion}
    \infPOVM[d_B] = \PP^{d_B}\infPOVM,    
\end{align}
where $\PP^{d_B} = \begin{pmatrix}
    1-d_B & 0\\
    d_B & 1
\end{pmatrix}$
is a stochastic matrix that describes the classical post-processing, and $\vec{\Gamma} = \begin{pmatrix}
    \Gamma_\nc\\
    \Gamma_\c
\end{pmatrix}.$
Similarly, for a more general detection setup with $k$ threshold detectors, we can write the post-processing map acting on all $2^k$ click patterns based on the dark count rates $\dB = (d_1, d_2, \cdots, d_k)$ as $\PPdb$. While the exact structure of $\PPdb$ is not important, we shall comment on the relevant properties of $\PPdb$ where applicable.
In this work, we will refer to the $(i,j)$ element of a post-processing matrix $\PP$ as $\PP{i}{j}$ to represent the probability of outcome (alternately, click pattern) $i$ given outcome (alternately, click pattern) $j$.

\subsection{Modelling detector efficiency}

Detector efficiency refers to a detector not always registering a click when photons are incident on it. In this work, we primarily consider the case where the detector efficiency is independent of prior detection events. However, practical detectors exhibit a finite recovery period following a detection event, commonly known as dead time. The effects of dead time are considered only in \cref{sec:corrDetEAT}, where we defer the development of a detailed model.

A typical quantum optical model for detection efficiency uses the single-photon detection efficiency $\eta$, which represents the probability of the detector registering a click for each incident photon.
This can also be represented as a beam splitter with transmission amplitude coefficient $\sqrt{\eta}$, followed by a lossless detector.

An important special case that we shall use in this work, is that of a lossy detector with no dark counts, where the input has \textit{no more than a single photon}. This case might sound oddly specific, but its usage will become clearer in \cref{subsec:noiseChannelLoss} (and in the proof of \cref{thm:lossNoiseChannel}). In this case, the loss can be represented as a classical post-processing analogous to \cref{eq:DCPPforSingleDetVecVersion}
\begin{align} \label{eq:lossAsAPP}
    \infPOVM[\eta] = \PP^{\eta}\infPOVM,
\end{align}
where $\PP^{\eta} = \begin{pmatrix}
    1 & 1-\eta\\
    0 & \eta
\end{pmatrix}$, and $\vec{\Gamma} = \begin{pmatrix}
    \Gamma_\nc\\
    \Gamma_\c
\end{pmatrix}.$ Similarly, for a more general detection setup with $k$ threshold detectors, we can write the post-processing map $\PPloss$ acting on all $2^k$ click patterns based on the loss of each detector $\loss = (\eta_1,\eta_2, \cdots, \eta_k)$. Note once again that this post-processing map is valid for detectors without dark counts, with an input with no more than a single photon.
We stress that at no point do we assume that our detectors only allow for a single photon input; we simply detailed the model in the single-photon subspace.

\begin{notation}
    Throughout this work, we use post-processing matrices to describe detection setups involving multiple detectors. The rows and columns of these matrices are ordered by increasing numbers of detection events.
\end{notation}

\subsection{Multi-mode detectors} \label{subsec:multimode}

The behaviour of realistic detectors across multiple spatial and temporal modes is not fully understood, and a complete physical model amenable to theoretical analysis remains an open problem. We therefore adopt a simplified model used in \cite{zhang_security_2021,tupkary2024phaseerrorrateestimation}. In this model, each mode is associated with its own efficiency and dark count rate. The overall measurement is then described as a block-diagonal combination of the single-mode POVM elements corresponding to each mode:
\begin{align}
    \Gamma^{\text{multi}}_i \;=\; \bigoplus_{\mathbf{d}} \Gamma_i\!\left(\eta_i(\mathbf{d}), d_i(\mathbf{d})\right),
\end{align}
where $\Gamma_i(\eta_i(\mathbf{d}), d_i(\mathbf{d}))$ denotes the single-mode POVM element corresponding to mode~$\mathbf{d}$.

\section{Squashing maps and noise channels} \label{sec:proofTechnique}

We first give a brief, intuitive overview of the core idea of our framework, which is an extension of squashing maps \cite{tsurumaru2008security,beaudry2008squashing,tsurumaru2010squash,gittsovich_squashing_2014}, before explaining the details.
For an arbitrary detection setup with dark count rates $\dB$ and loss $\loss$, we wish to construct a ``noise channel'' $\noiseChannel$ such that $\Tr[\infPOVM[\dB, \loss]\rho] = \Tr[\infPOVM \noiseChannel(\rho)]$ for all input quantum states $\rho$. Informally, we wish to model the noisy detection setup as a noise channel followed by an ideal detection setup.

The existence of such a channel is often sufficient to simplify the analysis for applications such as entanglement verification \cite{moroder_entanglement_2010} and QKD.
Intuitively, this is because the noise channel can be `given' to the adversary.
For example, in QKD the security of a protocol using noisy POVM $\infPOVM[\dB,\loss]$ would directly follow from the security of a protocol using ideal POVM $\infPOVM$ along with the existence of a noise channel (See Lemma 6 from Ref.~\cite{nahar_postselection_2024} for a proof).

Sometimes it is difficult to generically construct a noise channel acting on the full Fock space. Thus, to simplify our task we make use of squashing maps \cite{gittsovich_squashing_2014, zhang_security_2021}. The general framework, including the use of squashing maps is shown in \Cref{fig:setOfEquivalences}.
\begin{figure*}
    \centering
    \scalebox{0.8}{\begin{tikzpicture}
    \pic {detector={detinf, {$\infPOVM[\dB, \loss]$}, black}};

    \node[process, right of = detinf, xshift = 3cm, olive] (PP){$\Msq'$};
    
    \draw (-5,0) -- ([xshift=-0.0cm]detinf.west); 

    \draw[dashed] ([xshift=0.22cm]detinf.east) -- (PP);
    
    \begin{scope}[shift={(-1,-4)}]
        \pic {detector={detinfideal, $\infPOVM[\loss]$,black}};
        \node[process, right of = detinfideal, xshift = 1.5cm, orange](DCPP){$\PP^{\dB}$};
        \node[process, right of = DCPP, xshift = 1.5cm, olive] (SqPP){$\Msq'$};
        
        \draw (-4,0) -- ([xshift=-0.2cm]detinfideal.west);
        \draw[dashed] ([xshift=0.35cm]detinfideal.east) -- (DCPP);
        \draw[dashed] (DCPP.east) -- (SqPP);
    \end{scope}

    \begin{scope}[shift={(-1,-8)}]
        \pic {detector={detinfideal2, $\infPOVM[\loss]$,black}};
        \node[process, right of = detinfideal2, xshift = 1.5cm, orange](SqPPNew){$\Msq$};
        \node[process, right of = SqPPNew, xshift = 1.5cm, olive] (DCPPNew){$\Mdc$};
        
        \draw (-4,0) -- ([xshift=-0.2cm]detinfideal2.west);
        \draw[dashed] ([xshift=0.35cm]detinfideal2.east) -- (SqPPNew);
        \draw[dashed] (SqPPNew.east) -- (DCPPNew);
    \end{scope}

    \begin{scope}[shift={(-1,-11.5)}]
        \node[process](SqMap){$\Lambda$};
        \pic [right of = SqMap,xshift = 1cm]{detector={detfinideal, $\finPOVM[\loss]$, black}};
        \node[process, right of = detfinideal, xshift = 2cm, olive] (DCPPNew){$\Mdc$};
        
        \draw (-4,0) -- (SqMap.west);
        \draw (SqMap) -- ([xshift=-0.2cm]detfinideal.west);
        \draw[dashed] ([xshift=0.35cm]detfinideal.east) -- (DCPPNew);
    \end{scope}

    \begin{scope}[shift={(-1,-15)}]
        \node[process,xshift = -1cm](SqMap){$\Lambda$};
        \node[process,right of = SqMap,xshift = 2cm](NoiseChannel){$\noiseChannel$};
        \pic [right of = NoiseChannel,xshift = 2cm]{detector={detfinideal, $\finPOVM$,black}};
        
        \draw (-4,0) -- (SqMap.west);
        \draw (SqMap) -- (NoiseChannel);
        \draw (NoiseChannel) -- ([xshift=-0.25cm]detfinideal.west);
    \end{scope}
    
    \node at (0,-2) {\scalebox{2}{$\Big\Updownarrow$}};
    \node at (-0.5,-2) {(i)};
    
    \node at (0,-6) {\scalebox{2}{$\Big\Updownarrow$}};
    \node at (-0.5,-6) {(ii)};
    \node[text=red] at (0.5,-6) {\scalebox{2}?};

    \node at (0,-10) {\scalebox{2}{$\Big\Updownarrow$}};
    \node at (-0.5,-10) {(iii)};

    \node at (0,-13.4) {\scalebox{2}{$\Big\Updownarrow$}};
    \node at (-0.5,-13.4) {(iv)};
    \node[text=red] at (0.5,-13.4) {\scalebox{2}?};
\end{tikzpicture}}
    \caption{A set of equivalences (as quantum-to-classical measurement channels) of detection setups to consider dark counts as part of the noise channel. Here, $\Msq'$ represents the classical post-processing carried out in the protocol, $\PPdb$ is the dark count post-processing, $\Msq$ is the post-processing required for the existence of the squashing map $\Lambda$, and $\Mdc$ is the post-processing fixed by \cref{eq:PPswap}. In the figure, olive post-processing maps represent free choices that can be made to make equivalences (ii) and (iv) hold. Orange post-processing maps represent stochastic processes fixed by equivalences (i) and (iii).} \label{fig:setOfEquivalences}
\end{figure*}
Each equivalence is an equivalence of quantum-to-classical measurement channels.
\begin{itemize}
    \item We start by considering a generic detection setup that uses threshold detectors with dark counts and loss. Additionally, we allow for some classical post-processing $\Msq'$. We elaborate more on this choice later.
    \item Equivalence (i) follows from the fact that dark counts can be modelled as a classical post-processing $\PPdb$ as described in \Cref{subsec:DCPP}.
    \item Part of our task is to show that equivalence (ii) holds for some choice of $\Mdc$ and a fixed $\Msq$, as discussed in detail in \cref{subsubsec:swapPPCond}.
    \item Equivalence (iii) assumes that there exists a squashing model \cite{gittsovich_squashing_2014,zhang_security_2021} for the particular detection setup being considered together with the specific $\Msq$ fixed through equivalence (ii). Although, the simple squashers described in Ref.~\cite{gittsovich_squashing_2014} have been shown to exist only under some restrictions on the detection setup, the flag-state squasher \cite{zhang_security_2021} and the weight-preserving flag-state squasher (WPFSS) \cite[Lemma 7]{nahar_postselection_2024} always exist for detection setups using threshold detectors.
    \item It is then left to show that equivalence (iv) holds, which is discussed in detail in \cref{subsubsec:noiseChannelCond}.
\end{itemize}
    
\subsubsection{Does equivalence (ii) hold?} \label{subsubsec:swapPPCond}

$\Msq$ is the post-processing required for the specific squashing map we wish to use. For example, for an active BB84 detection setup, we can use Theorem 10 from Ref.~\cite{gittsovich_squashing_2014} to obtain the post-processing $\Msq = \begin{pmatrix}
                        1&0&0&0\\
                        0&1&0&1/2\\
                        0&0&1&1/2
                    \end{pmatrix}$
which maps double clicks in a particular basis to random single clicks in the same basis.
$\PPdb$ is the dark count post-processing for the detection setup. Thus, the equivalence amounts to asking if there exist stochastic matrices $\Msq'$, $\Mdc$ such that
\begin{align} \label{eq:PPswap}
    \Msq' \PPdb = \Mdc\Msq.
\end{align}

Note that $\Msq'$ can be freely chosen by physically performing the appropriate classical post-processing on the measurement results. For simplicity, we choose $\Msq' = \Msq$ in this work. We shall comment more on the post-processing $\Msq$ needed for the flag-state squasher and the WPFSS in \cref{sec:appToPS}.

\subsubsection{Does equivalence (iv) hold?} \label{subsubsec:noiseChannelCond}

This equivalence can be formally stated as
\begin{align} \label{eq:noiseChannelCond}
    \Mdc\Tr[\finPOVM[\loss] \rho] = \Tr[\finPOVM\ \noiseChannel(\rho)],
\end{align}
for all density matrices $\rho$ in the squashed space. Here, $\finPOVM[\loss]$ is the squashed lossy POVM which satisfies $\Lambda^\dag(\finPOVM[\loss]) = \infPOVM[\loss]$, and $\finPOVM$ is the ideal target POVM.

Thus, we have a framework that allows us to reduce the analysis of a detection setup with dark counts and loss $\infPOVM[\dB, 
\loss]$ to that of a squashed ideal POVM $\finPOVM$. This reduction follows simply by checking if \cref{eq:PPswap,eq:noiseChannelCond} hold. For completeness, in \cref{app:numCheck} we give a general method to numerically check if \cref{eq:PPswap,eq:noiseChannelCond} hold, given a specific value of $\dB$ and $\loss$.

\subsection{Pedagogical example - Active BB84 detection setup with lossless detectors} \label{subsec:pedExamp}

We illustrate the use of our framework via the concrete example of the active BB84 detection setup, using the simple squasher described in \cite[Theorem 10]{gittsovich_squashing_2014}. For the purpose of this example, we assume that both detectors have the same loss. This loss can be given to Eve \cite[Fig. 1]{koashi2006efficient}, and so is equivalent to assuming lossless detectors.

We first analyse equivalence (ii) for each basis choice separately. The dark count post-processing for this setup is $\PPdb = \begin{pmatrix}
    (1-d_1)(1-d_2) & 0  & 0 & 0\\
    d_1(1-d_2) & (1-d_2) & 0 & 0\\
    d_2(1-d_1) & 0 & (1-d_1) & 0\\
    d_1d_2 & d_2 & d_1 & 1
\end{pmatrix}.$ Also, recall that the squashing post-processing for this protocol is $\Msq = \begin{pmatrix}
    1&0&0&0\\
    0&1&0&1/2\\
    0&0&1&1/2
\end{pmatrix}.$

We make the simplifying choice $\Msq' = \Msq$, and compute
\begin{align} \label{eq:PsqPdb}
    \Msq \PPdb=
     \begin{pmatrix}
            (1-d_1)(1-d_2) & 0  & 0 & 0\\
            d_1(1-d_2/2) & 1-d_2/2 & d_1/2 & 1/2\\
            d_2(1-d_1/2) & d_2/2 & 1-d_1/2 & 1/2
        \end{pmatrix}.
\end{align}
We then attempt to construct $\Mdc$ so that it satisfies \cref{eq:PPswap}. First, we can write $\Msq = \begin{pmatrix}
        \mathbb{I}_3 & v
\end{pmatrix},$ where $v = \begin{pmatrix}
                                0\\
                                1/2\\
                                1/2
                            \end{pmatrix}.$ Then,
\begin{align} \label{eq:PdcPsq}
    \Mdc \Msq = \begin{pmatrix}
                    \Mdc & u
                \end{pmatrix},
\end{align}
where $u = \Mdc v$. As can be seen from \cref{eq:PsqPdb,eq:PdcPsq}, \cref{eq:PPswap} can only hold if
\begin{align}
\label{eq:DCPPafterSwap}
    \Mdc = \begin{pmatrix}
        (1-d_1)(1-d_2) & 0  & 0\\
        d_1(1-d_2/2) & 1-d_2/2 & d_1/2\\
        d_2(1-d_1/2) & d_2/2 & 1-d_1/2\\
    \end{pmatrix},
\end{align}
which in turn implies that
\begin{align}
    u = \begin{pmatrix}
        0\\
        1/2 + \frac{d_1-d_2}{2}\\
        1/2 + \frac{d_2-d_1}{2}
    \end{pmatrix}.
\end{align}
Note that this does \emph{not} satisfy \cref{eq:PPswap} unless $d_1 = d_2 \eqqcolon d$.  Thus, we must assume that the dark count rates of both detectors are exactly the same to use this framework with the qubit squasher.

Although this assumption is not met in practice, we nonetheless elaborate on this example due to its pedagogical value in illustrating our general framework.
In \cref{sec:NoiseChannelWithFSS} we shall avoid this assumption by using the flag-state squasher instead.
    
Moving our attention to the construction of the noise channel described in equivalence (iv), we construct the map
\begin{align}
    \nonumber \noiseChannel{\dB}(\rho) = &\Tr[\rho\ketbra{\vac}]\;
    \left((1-d)^2\;\ketbra{\vac}+ d(1-d/2)\;\Pi\right)\\
    &+ (1-d)\; \Pi\rho\Pi + \Tr[\Pi \rho \Pi]\,d\;\frac{\Pi}{2},\label{eq:NoiseChannelSimpleConstruction}
\end{align}
where $\Pi$ is the projection onto the qubit space. It can easily be verified that this guess is a valid channel by writing it as a QND measurement, followed by state preparation and depolarizing channels.

Finally, explicit computations give us that \cref{eq:noiseChannelCond} is satisfied by this channel. We sketch the computations for a single basis here.
First consider the POVM elements $F_{\text{nc}} = \ketbra{\vac}$, $F_0 = \ketbra{0}$, and $F_1 = \ketbra{1}$. From \cref{eq:DCPPafterSwap}, we obtain
\begin{align} \label{eq:BB84egLHSNoiseChannelEq}
	\Mdc \Tr[\finPOVM \rho] &= \begin{pmatrix}
									(1-d)^2 \Tr[\rho \ketbra{\vac}]\\
                					d(1-d/2) \Tr[\rho \ketbra{\vac}] + (1-d/2) \Tr[\rho\ketbra{0}] + d/2\Tr[\rho\ketbra{1}]\\
                					d(1-d/2) \Tr[\rho \ketbra{\vac}] + d/2\Tr[\rho\ketbra{0}] + (1-d/2) \Tr[\rho\ketbra{1}]
            					\end{pmatrix}.
\end{align}
The claim $\Tr[\finPOVM\;\noiseChannel{\dB}(\rho)] = \Mdc \Tr[\finPOVM\rho]$ then follows from \cref{eq:NoiseChannelSimpleConstruction,eq:BB84egLHSNoiseChannelEq} by noting that $\Pi\ket{0} = \ket{0}$ and $\Pi\ket{1} = \ket{1}$.
A similar computation for the $+/-$ basis gives us that \cref{eq:noiseChannelCond} is satisfied for this noise channel.

Thus, we have shown by explicit construction, the existence of a noise channel for the active BB84 setup using a simple squasher \cite[Theorem 10]{gittsovich_squashing_2014}, under the assumption that both the lossless detectors have the same dark count rate. 
Similar claims have appeared in the literature (e.g., \cite{lydersenSecurity2010}), but without a formal proof. The reliance on numerous unphysical assumptions in our construction highlights the non-trivial nature of the task.
We now extend our analysis to the more practical scenario in which the detectors may have differing loss and dark count rates, using the flag-state squasher to facilitate this generalisation.

\section{Noise channel with flag-state squasher} \label{sec:NoiseChannelWithFSS}

In this section, we introduce the flag-state squasher \cite{zhang_security_2021} in \cref{subsec:FSS} to generalise the pedagogical example described in \cref{subsec:pedExamp} to the scenario with unequal loss and dark count rates (\cref{subsec:noiseChannelDarkCounts,subsec:noiseChannelLoss}).
Furthermore, we extend our analysis to more generic setups in \cref{subsubsec:NoiseChannelGenSetups}.

\subsection{Flag-state squasher} \label{subsec:FSS}

The flag-state squasher \cite[Theorem 1]{zhang_security_2021} is a squashing map (equivalence (iii) in \cref{fig:setOfEquivalences}) that generalises the simple squasher \cite{beaudry2008squashing,gittsovich_squashing_2014} for arbitrary detection setups where the POVM elements have a block-diagonal structure. For detection setups using threshold detectors, such a block-diagonal structure arises naturally where the blocks correspond to total photon number:
\begin{align*}
    \Gamma_i = \Gamma_{i,0}\oplus \Gamma_{i,0<m\leq \nFSS} \oplus \Gamma_{i,m > \nFSS}.
\end{align*}
Here, $m > \nFSS$ corresponds to the set of photon-numbers $m$ greater than the cutoff $\nFSS$, and $\Gamma_{i,0}$ acts on the space spanned by the vacuum state $\ket{\vac}$. Although separating out the vacuum space does not affect the squashing --- the space spanned by the states with photon number $m\leq \nFSS$ can correspond to a single block --- we explicitly separate the vacuum space as it simplifies the notation later.
The target measurements are given by
\begin{align*} 
    F_i = \Gamma_{i,0}\oplus \Gamma_{i,0<m\leq \nFSS} \oplus \ketbra{i},
\end{align*}
where $\{\ket{i}\}_{i=0}^{\nmeas-1}$ forms an orthonormal set of vectors termed `flags'. We denote the space spanned by $\{\Gamma_{i,0<m\leq \nFSS}\}_{i=1}^{\nmeas}$ as $\H_{0<m\leq\nFSS}$, the space spanned by the vacuum state as $\H_0$, and the space spanned by the flags as $\H_{F}$. We call $\H_0\oplus \H_{0<m\leq\nFSS}$ the preserved subspace, and $\H_{F}$ the flag space. Importantly, note that the portion of the target measurements acting on the preserved subspace is identical to that of the actual measurement.

Note that in adversarial applications, the idea is to `give' the eavesdropper Eve the squashing map so that the analysis can then be restricted to the finite-dimensional POVM elements. However, in the case of the flag-state squasher the existence of the flags implies that there exists a classical state\footnote{By classical we mean diagonal in the basis described by the flags.} living entirely in the flag subspace that results in a given probability distribution when measured, for all probability distributions.
Thus, giving Eve full control over the flag-state squasher would result in a complete loss of any `quantumness' of the quantum information protocol (since any observations can be completely explained by classical states).
A common solution is to add an additional constraint that bounds the weight $W$ in the flag space, intuitively bounding the extent to which a classical flag state can explain observations.

The canonical method of bounding the weight outside the preserved subspace can be found in Refs.~\cite{narasimhachar2011study,li_application_2020} and proceeds as follows. For any event $e$\footnote{We use the term event to refer to an outcome, or a set of outcomes, of a POVM acting on a single round of the protocol. This should not be confused with higher-level events such as accept or abort decisions made during the course of a QKD protocol. Throughout this work, we consistently use event in the former sense and do not consider the latter.}, and any input state $\rho$ it can be shown that
\begin{align} \label{eq:wtOutsideSubspaceGeneric}
    1-\Tr[\rho \Pi_{\nFSS}] \leq \frac{p(e)- \lambdamin{\Pi_{\nFSS}\Gamma_e\Pi_{\nFSS}}}{\lambdamin{\overline{\Pi}_{\nFSS}\Gamma_e\overline{\Pi}_{\nFSS}}-\lambdamin{\Pi_{\nFSS}\Gamma_e\Pi_{\nFSS}}},
\end{align}
where $\Pi_{\nFSS}$ $(\overline{\Pi}_{\nFSS})$ is the projection on (outside) the space corresponding to $\Gamma_{i,m\leq \nFSS}$. Note that the choice of this event $e$ need not be a single POVM element (alternately, click-pattern), it could also be some `coarse-graining' that includes multiple click-patterns. Thus, there is a large amount of freedom when choosing this event.

When working with the infinite-dimensional POVM, some protocol-dependent choices \cite{li_improving_2020,nahar_imperfect_2023} for the event $e$ lead to good bounds on the weight $W$. This bound can then be added in as an additional constraint to the finite-dimensional analysis. More recently, Refs.~\cite{kamin2024improved,wang2025phase} show that this event $e$ can be chosen to be the multi-click event, i.e. all click patterns that consist of more than a single click in the various detectors, for arbitrary passive optical setups. This is the choice we shall focus on in \cref{sec:appToPS} due to its generality.

Thus, our task can be broken up into two parts as shown in \cref{fig:FSSEquivalences}.
\begin{figure*}
    \centering
    \scalebox{0.8}{\begin{tikzpicture}

\begin{scope}[shift = {(1,0)}]
    \pic {detector={detinf, {$\infPOVM[\dB, \loss]$}, black}};
    
    \draw (-5,0) -- ([xshift=-0.05cm]detinf.west); 
\end{scope}
        
    \begin{scope}[shift={(-1,-4)}]
        \pic {detector={detinfideal, $\infPOVM[\loss]$,black}};
        \node[process,right of = detinfideal, xshift = 1.5cm](DCPP){$\PP^{\dB}$};
        
        \draw (-3,0) -- ([xshift=-0.25cm]detinfideal.west);
        \draw[dashed] ([xshift=0.35cm]detinfideal.east) -- (DCPP);
    \end{scope}

    \begin{scope}[shift={(-1,-8)}]
        \node[process](SqMap){$\Lambda_{\text{FSS}}$};
        \pic [right of = SqMap,xshift = 1cm]{detector={detfinideal, $\finPOVM[\loss]$\\$W$,black}};
        \node[process,right of = detfinideal, xshift = 2cm] (DCPPNew){$\PP^{\dB}$};
        
        \draw (-3,0) -- (SqMap.west);
        \draw (SqMap) -- ([xshift=-0.25cm]detfinideal.west);
        \draw[dashed] ([xshift=0.35cm]detfinideal.east) -- (DCPPNew);
    \end{scope}

    \begin{scope}[shift={(-1,-12)}]
        \node[process,xshift = -1cm](SqMap){$\Lambda_{\text{FSS}}$};
        \node[process,right of = SqMap,xshift = 2cm](NoiseChannel){$\noiseChannel$};
        \pic [right of = NoiseChannel,xshift = 2cm]{detector={detfinideal, $\finPOVM$\\$W^{\dB, \loss}$,black}};
        
        \draw (-3,0) -- (SqMap.west);
        \draw (SqMap) -- (NoiseChannel);
        \draw (NoiseChannel) -- (detfinideal.west);
    \end{scope}
    
    \node at (0,-2) {\scalebox{2}{$\Big\Updownarrow$}};
    \node at (-0.5,-2) {(i)};
    
    \node at (0,-6) {\scalebox{2}{$\Big\Updownarrow$}};
    \node at (-0.5,-6) {(ii)};

    \node at (0,-10) {\scalebox{2}{$\Big\Updownarrow$}};
    \node at (-0.55,-10) {(iii)};
    \node[text=red] at (0.5,-10) {\scalebox{2}?};
\end{tikzpicture}}
    \caption{A set of equivalences (as quantum-to-classical measurement channels) of detection setups to consider dark counts and loss as part of the noise channel, using the flag-state squasher.} \label{fig:FSSEquivalences}
\end{figure*}
First, we need to show that there exists a noise channel $\noiseChannel$ such that \cref{eq:noiseChannelCond} is satisfied. Next, we need to find the weight outside the preserved space $W^{\dB, \loss}$ after the application of the noise channel. In other words, given that $W\geq 1-\Tr[\rho\Pi_\nFSS]$ for any input state $\rho$, we need to find $W^{\dB, \loss} \geq 1-\Tr[\noiseChannel(\rho)\Pi_\nFSS]$.

Note that unlike the qubit squasher, the flag-state squasher does not depend on some post-processing $\Msq$ for its existence. Thus, \cref{fig:FSSEquivalences} does not contain equivalence (ii) from \cref{fig:setOfEquivalences}. However, we might sometimes wish to `coarse-grain' the data before using the flag-state squasher. In this case, we would need to also prove equivalence (ii) (from \cref{fig:setOfEquivalences}) for the coarse-graining post-processing. However, for the sake of pedagogy, we will first perform our analysis without any such complications before discussing the impact of coarse-graining in \cref{sec:appToPS}.

\subsection{Construction of Noise Channel} \label{subsec:noiseChannelConstructions}

We shall construct the full noise channel $\noiseChannel$ in two steps. First we construct a noise channel for the dark counts $\noiseChannel{\dB}$, and then another for the loss $\noiseChannel{\loss}$.

\subsubsection{Noise channel for dark counts}
\label{subsec:noiseChannelDarkCounts}

Any detection setup has single-click events, defined as events where only a single detector clicks, and multi-click events, defined as events where more than one detector clicks.
We let $\singleSet$ be the set of single-click events, and $\multiSet$ be the set of multi-click events. The no-click event will always be denoted by the event $0$. The set of all events will be denoted by $U$.

We first list a set of sufficient conditions we require from the classical post-processing $\PPdb$ in order to construct the noise channel. We then physically motivate these conditions.
\begin{enumerate}
    \item \label{step:sIsS}The post-processing does not turn one single-click event into another, i.e., 
    \begin{align} \label{eq:sIsS}
        \PPdb{s}{s'} = 0,
    \end{align}
    for all distinct $s$, $s' \in \singleSet$.
    \item The post-processing does not turn any click event into a no-click event, i.e.,
    \begin{align}\label{eq:noWhiteCounts}
        \PPdb{0}{i} = 0,
    \end{align}
    for all $i\neq 0$.
    \item The post-processing is such that
    \begin{align} \label{eq:11morethan00}
        \PPdb{s}{s} \geq \PPdb{0}{0},
    \end{align}
    for all $ s\in\singleSet$.
\end{enumerate}
The first two conditions are true for any dark count post-processing as dark counts do not stop a detector from clicking. To see why the third condition is true, observe that $\PPdb{s}{s} = \prod_{i\neq s} (1-d_i)$ is the probability that none of the other detectors have a dark count. On the other hand, $\PPdb{0}{0} = \prod_{i} (1-d_i)$ is the probability that none of the detectors have a dark count. Since $(1-d_s)\leq 1$, the claim follows.

We also need to make an additional assumption on the POVM without dark counts $\infPOVM[\loss]$. Informally, the assumption is that we cannot have more clicks than photons. More formally, this can be expressed as
\begin{equation} \label{eq:noSingleToMulti}
    \begin{aligned}
        \Tr[\Gamma^{\loss}_{m} \rho_1] &= 0\\
        \Tr[\Gamma^{\loss}_{m} \ketbra{\vac}] &= 0\\
        \Tr[\Gamma^{\loss}_{s} \ketbra{\vac}] &= 0,
    \end{aligned}
\end{equation}
for all $m \in \multiSet$, $s \in \singleSet$ and any single-photon state $\rho_1$. This naturally holds for typical detection setups as dark counts are the only way single photons can result in multiple detectors clicking or vacuum can result in any detector clicking. Note that this immediately implies that $\Tr[\Gamma_0\ketbra{\vac}] = 1$.

We combine all these properties to define a generic optical detection setup.
\begin{definition}[Threshold detection setup with independent dark counts] \label{def:detWithDarkCounts}
    We define the POVM $\infPOVM[\dB, \loss]$ to be a \emph{threshold detection setup with independent dark counts} if
    \begin{enumerate}
        \item it is equivalent to a POVM without dark counts $\infPOVM[\loss]$ followed by a dark count post-processing $\PPdb$ as described in equivalence (i) of \cref{fig:FSSEquivalences},
        \item the dark count post-processing satisfies \cref{eq:11morethan00,eq:noWhiteCounts,eq:sIsS}, and
        \item the POVM without dark counts $\infPOVM[\loss]$ satisfies \cref{eq:noSingleToMulti}.
    \end{enumerate}
\end{definition}

We now have all the tools to construct the noise channel for dark counts.
\begin{restatable}[Dark count noise channel]{theorem}{FSSFineGrained}\label{thm:FSSFineGrained}
    Let $\infPOVM[\dB, \loss]$ be the POVM for a threshold detection setup with independent dark counts, where each $\Gamma^{\dB, \loss}_i = \Gamma_{i,0}^{\dB, \loss}\oplus\Gamma^{\dB, \loss}_{i,m=1} \oplus  \Gamma^{\dB, \loss}_{i,m > 1}$ is block-diagonal with an associated Hilbert space $\H_0\oplus\H_{1}\oplus \H_{m>1}$ corresponding to the total photon number. Then there exists a flag-state squashing map $\Lambda: \H_0\oplus\H_{1}\oplus \H_{m>1} \xrightarrow[]{} \H_{m\leq 1}\oplus\H_F$, and a noise channel $\noiseChannel{\dB}: \H_0\oplus\H_{1}\oplus\H_F\xrightarrow[]{} \H_0\oplus\H_{1}\oplus\H_F$ such that $\Tr[\infPOVM[\dB, \loss]\rho] = \Tr[\finPOVM[\loss]\ \noiseChannel{\dB}\left(\Lambda(\rho)\right)]$ for all density matrices $\rho$. Here,
    \begin{align*} 
        F^\loss_i = \Gamma^\loss_{i,m=0} \oplus\Gamma^\loss_{i,m=1}\oplus  \ketbra{i},
    \end{align*}
    where $\ket{i}$ forms an orthonormal basis for the flag space $\H_F$. Moreover for any density matrix $\rho$,
    \begin{align} \label{eq:subspaceWtIncrease}
        \Tr[\noiseChannel{\dB}(\rho)\Pi_{\leq 1}] = \PP{0}{0}^\dB\Tr[\rho\Pi_{\leq 1}],
    \end{align}
    where $\Pi_{\leq 1}$ is the projection onto the space $\H_0\oplus\H_1$, and $\PPdb$ is the stochastic matrix that models the dark counts.
\end{restatable}
The proof is given in \cref{app:noiseChannelProofs}.
\begin{multimode}
    Note that the multi-mode detector model considered in \cref{subsec:multimode} also fits the conditions in \cref{def:detWithDarkCounts}. Thus, \cref{thm:FSSFineGrained} can also be used for these multi-mode detectors. Moreover, we believe that these conditions are general enough that any physical model of multi-mode detectors should satisfy them.
\end{multimode}

Recall that equivalence (iii) in \cref{fig:FSSEquivalences} required the existence of a noise channel $\noiseChannel$, and a bound on the weight outside the preserved subspace $W^{\dB, \loss}$ after the application of the noise channel.
\cref{thm:FSSFineGrained} is the first step towards equivalence (iii), showing that there exists a noise channel $\noiseChannel{\dB}$ for dark counts, and relating the weight outside the preserved subspace before and after the application of the dark count noise channel through \cref{eq:subspaceWtIncrease} --- $W^\dB = 1-\PPdb{0}{0}(1-W)$. We now prove an analogous result for lossy detectors.

\subsubsection{Noise channel for loss} \label{subsec:noiseChannelLoss}

\begin{restatable}[Loss noise channel]{theorem}{lossNoiseChannel}\label{thm:lossNoiseChannel}
    Let $\infPOVM[\loss]$ be the POVM for a threshold detection setup with loss (and without dark counts), where for each POVM outcome $i$, $\Gamma^{\loss}_i = \Gamma_{i,m=0}\oplus\Gamma^{\loss}_{i,m= 1} \oplus  \Gamma^{\loss}_{i,m > 1}$ is block-diagonal with an associated Hilbert space $\H_0\oplus\H_{1}\oplus \H_{m>1}$ corresponding to the total photon number. Here, $\Gamma_{i,m=0}$ is independent of loss $\loss$ for all $i$.
    Let $\etamin$ and $\etamax$ be the minimum and maximum of all the detector efficiencies used in the detection setup. We can then define a family of noise channels $\noiseChannel{\loss}^{\idealEff}$ and target POVMs $\finPOVM[\idealEff]$ parametrised by some parameter $\idealEff$ as follows.
    For any value of $\idealEff \in \left[\frac{\etamin}{1-(\etamax-\etamin)},1\right]$,
    there exists a flag-state squashing map $\Lambda: \H_0\oplus\H_{1}\oplus \H_{m>1} \xrightarrow[]{} \H_0\oplus\H_{1}\oplus\H_F$, and a noise channel $\noiseChannel{\loss}^{\idealEff}: \H_0\oplus\H_{1}\oplus\H_F\xrightarrow[]{} \H_0\oplus\H_{1}\oplus\H_F$ such that $\Tr[\infPOVM[\loss]\rho] = \Tr[\finPOVM[\idealEff]\ \Phi_{\loss}^{\idealEff}\left(\Lambda(\rho)\right)]$ for all density matrices $\rho$. Here,
    \begin{align*} 
        F^{\idealEff}_i = \Gamma_{i,m=0}\oplus\Gamma^{\idealEff}_{i,m=1} \oplus  \ketbra{i},
    \end{align*}
    where $\ket{i}$ forms an orthonormal basis for the flag space $\H_F$, and $\infPOVM[\idealEff]$ is the POVM with (common) efficiency $\idealEff$ in all detectors.
    Moreover for any density matrix $\rho$,
    \begin{align} \label{eq:subspaceWtIncreaseLoss}
        \Tr[\noiseChannel{\loss}^{\idealEff}(\rho)\Pi_{\leq 1}] = \Tr[\rho\Pi_0]+\frac{\etamin}{\idealEff} \Tr[\rho\Pi_{1}],
    \end{align}
    where $\Pi_{1}$ is the projection onto the space $\H_{1}$, $\Pi_{0}$ is the projection onto the space $\H_{0}$, and $\Pi_{\leq1}$ is the projection onto the space $\H_0\oplus\H_1$.
\end{restatable}
The proof is given in \cref{app:noiseChannelProofs}.

Note that \cref{thm:lossNoiseChannel} leaves us with some freedom to choose how we define the `ideal' POVM $\finPOVM[\idealEff]$. Increasing the efficiency $\idealEff$ of the ideal POVM decreases the weight $\frac{\etamin}{\idealEff}$ of the state in the preserved subspace.
Additional work must be done to decide what the optimal choice would be. Intuitively, we would expect that the optimal choice would maximise the weight in the preserved subspace, as any part of the state outside the preserved subspace is entirely classical. 

In practice, we often do not know the exact value of $\etamin$ and $\etamax$, but only have bounds on these values. These bounds can straightforwardly be used to obtain a bound on the weight described in \cref{eq:subspaceWtIncreaseLoss}. With regards to the common efficiency $\idealEff$, we note that for linear optical detection setups, the common efficiency can be considered to be a part of the channel\footnote{for e.g. as described in \cite[Section IV. B.]{lutkenhaus1999estimates}, \cite[Section III C]{zhang_entanglement_2017} or \cite[Fig. 1]{koashi2006efficient} for 
a polarisation-encoded BB84 setup}. Thus, the ``worst case'' common efficiency can be taken to be $1$.
\begin{multimode}
    The main property used in the proof of \cref{thm:lossNoiseChannel} was the model for lossy detectors in the $m\leq 1$ photon subspace (\cref{eq:lossAsAPP}).
    Thus, the results of \cref{thm:lossNoiseChannel} directly carry over to the multi-mode model described in \cref{subsec:multimode}. In this case, the noise channel $\noiseChannel{\loss}^{\idealEff}$ can be constructed to be a direct sum of noise channels for each mode, as \cref{eq:lossAsAPP} would hold independently for each mode.
    More detailed models would need to be dealt on a case-by-case basis, or alternately via the more general \cref{thm:genericNoiseChannel} described below.
\end{multimode}

We now refer to the noise channel $\noiseChannel{\loss}^{\idealEff}$ as $\noiseChannel{\loss}$ for notational simplicity.
Finally, combining \cref{thm:FSSFineGrained,thm:lossNoiseChannel} gives us the noise channel $\noiseChannel = \noiseChannel{\loss}\circ\noiseChannel{\dB}$ required for equivalence (iii). The above theorems additionally relate the weight outside the preserved subspace before and after the application of the dark count noise channel through \cref{eq:subspaceWtIncrease,eq:subspaceWtIncreaseLoss}.

Note that the usage of this theorem towards equivalence (iii) still requires a bound on the weight outside the preserved subspace $W$ before the application of the noise channel.
Such a bound can be obtained for passive linear optical setups, as described in \cite[Section III]{wang2025phase}.
For \cref{thm:lossNoiseChannel} to yield tight bounds on the weight in the flag space, one must additionally upper bound the weight in the single-photon subspace following the action of the dark count noise channel. Such a bound is typically straightforward to compute; see \cref{subsec:EATActiveBB84} for a worked example that upper bounds the weight outside the $0$-photon subspace.
\begin{multimode}
    Even after using the noise channel to reduce the analysis to that of a multi-mode detector without loss and dark counts, the POVM elements describing this detector still have large dimensions due to the large number of modes. However, if we consider the simple model described in \cref{subsec:multimode}, the POVM elements in each mode are identical. Intuitively, this suggests that the overall analysis should further reduce to that of a single-mode POVM. While this can be shown to be true in a straightforward manner, we do not carry out a formal reduction here.
\end{multimode}

\subsubsection{Noise channel for generic setup imperfections} \label{subsubsec:NoiseChannelGenSetups}

We will now state a theorem that accounts for more generic detection setup imperfections, inspired by the proof of \cref{thm:FSSFineGrained}. Intuitively, the proof first decomposes the noisy POVM into an ideal POVM (with some probability $p$), and a non-ideal POVM (with probability $1-p$). The non-ideal POVM is then `turned' into a flag through the noise channel, leaving only the analysis of the ideal POVM.
This intuitive idea is formalised in the following theorem.

\begin{restatable}[Generic noise channel]{theorem}{genericNoiseChannel}\label{thm:genericNoiseChannel}
    Let the noisy POVM $\infPOVM[\noise]$ describing the actual imperfect detection setup and the noiseless POVM $\infPOVM[\ideal]$ have the same block-diagonal structure
    \begin{align*}
        \Gamma_i^\noise &=\bigoplus_{m} \Gamma_{i,\,m}^{\noise}\\
        \Gamma_i^\ideal &=\bigoplus_{m} \Gamma_{i,\,m}^{\ideal},
    \end{align*}
    for all POVM elements $i$ with associated Hilbert space $\mathcal{H}_B = \bigoplus_m \mathcal{H}_m$. Further, let the relation between each block of the noiseless and noisy POVM be given by
    \begin{align} \label{eq:deviationDefBlock}
        \infPOVM[m,\ \noise] - (1-q_m) \infPOVM[m,\ \ideal] \geq 0,
    \end{align}
    for $q_m\in [0,1]$ and all blocks $m$. Define an ideal flag POVM $\finPOVM[\ideal]$ as
    \begin{align} \label{eq:addedFlags}
        F_i^\ideal = \Gamma_i^\ideal \oplus \ketbra{i},
    \end{align}
    with associated Hilbert space $\H_{B}\oplus\H_F = \bigoplus_m\H_{m}\oplus\H_F$, and where $\{\ket{i}\}_{i=1}^{\nmeas}$ is an orthonormal basis for the flag space $\mathcal{H}_F$.
    Then there exists a noise channel $\noiseChannel{\noise}:\mathcal{H}_B\xrightarrow[]{}\H_B\oplus\H_F$ such that the following equations hold:
    \begin{align}
        \label{eq:genNoiseChannelConditionBlock} \Gamma_{i}^\noise &= \noiseChannel{\noise}^\dag\left[F_i^\ideal\right] \quad \forall\, i\\
        \label{eq:subspaceWtIncreaseGenericBlock} \noiseChannel{\noise}^\dag[\Pi_{m}] &= (1-q_m)\Pi_{m} \quad \forall\, m,
    \end{align}
    where $\Pi_{m}$ is the projection onto the block $\H_{m}$.
\end{restatable}
Unlike the other theorems, we leave the proof of this theorem in the main text as we believe the proof is simple to understand and instructive.
\begin{proof}
    First, note that \cref{eq:deviationDefBlock} implies that there exists a POVM $\vec{Q}_m$ such that
    \begin{align*}
        \infPOVM[m,\ \noise] = (1-q_m) \infPOVM[m,\ \ideal] + q_m \vec{Q_m}.
    \end{align*}
    We use this POVM to construct the noise channel for each block as depicted in \cref{fig:noiseChannelGeneric}:
    \begin{align}
        \noiseChannel{\noise}(\rho) \coloneqq \sum_m \left((1-q_m)\; \Pi_{m}\,\rho\,\Pi_{m}\ \oplus q_m \left(\sum_i \Tr\left[\Pi_{m}\,\rho\,\Pi_{m} Q_i\right] \ketbra{i} \right)\right),
    \end{align}
    where $\ketbra{i}$ are the flag states corresponding to measurement outcome $i$.
    \begin{figure*}[h]
        \centering
        \scalebox{0.95}{\begin{tikzpicture}[scale = 1.25]

\def\mylinewidth{2pt}

\coordinate (h1) at (4,0);
\coordinate (h2) at (10,0);
\coordinate (h3) at (0,-1);
\coordinate (h4) at (2,-1);
\coordinate (h5) at (4,-2);
\coordinate (h6) at (7,-2);
\coordinate (h8) at (10,-2);
\pic [right of = h5,xshift = 0.5cm,scale = 0.75]{detector={meas, $\vec{Q}$,black}};

\node[left of = h3, font=\LARGE](H1){$\H_{m}$};
\node[right of = h2, font=\LARGE](H1end){$\H_{m}$};
\node[right of = h8,xshift = 0 cm, font=\LARGE](H0HFend){$\H_F$};
\node[below of = H1end,yshift = -0.325cm, font = \LARGE]{$\oplus$};

\node[fit=(h1)(h4)(h5)(h6), draw, line width = \mylinewidth*2, inner sep=30pt](fitrect) {};

\draw[cyan, line width=\mylinewidth] (h1) -- (h2);
\draw[cyan, line width=\mylinewidth] (h3) -- (h4);
\draw[cyan, line width=\mylinewidth] (h5) -- ([xshift=-0.08cm]meas.west);
\draw[dashed, cyan, line width=\mylinewidth] ([xshift=0.16cm]meas.east) -- ([xshift = -0.08 cm]h8);

\draw[cyan, line width=\mylinewidth] (h4) -- (h1) node[midway, above, yshift = 0.2cm, black]{\large$1-q$};
\draw[cyan, line width=\mylinewidth] (h4) -- (h5) node[midway, below, yshift = -0.3cm, black]{\large$q$};

\node[above=0.5cm of fitrect, font=\LARGE] (Title) {\textbf{Noise Channel $\noiseChannel{m,\,\noise}$}};

\end{tikzpicture}}
        \caption{Constructive description of each block of the noise channel. Each line corresponds to a subspace of the input Hilbert space associated with the block-diagonal decomposition of the POVM elements. The dashed lines refer to classical states. The full noise channel is obtained by stacking each of these blocks together. $\H_F$ refers to the flag space, which is common for all the blocks $\H_m$.} \label{fig:noiseChannelGeneric}
    \end{figure*}
    It is straightforward to verify that \cref{eq:genNoiseChannelConditionBlock,eq:subspaceWtIncreaseGenericBlock} hold for this noise channel.
\end{proof}

\cref{thm:genericNoiseChannel} has a number of noteworthy features as listed below.
\begin{enumerate}
    \item \textbf{Noisy POVM $\infPOVM[\noise]$:} We have generically termed this a noisy POVM. This could be noisy either due to imperfection characterisation, or due to Eve having some limited control over them. Both are treated equivalently in our framework.
    \item \textbf{Ideal POVM $\finPOVM[\ideal]$:} There is a large amount of freedom in choosing this `ideal' POVM. The choice of common efficiency $\idealEff$ in \cref{thm:lossNoiseChannel} was a manifestation of this degree of freedom. There are two competing factors when making this choice of POVM. First the ideal POVM should be useful for the application being analysed. Second, the choice of ideal POVM would influence how `close' the noisy POVM is to the ideal POVM, and thus the `cost' of using \cref{thm:genericNoiseChannel} through \cref{eq:subspaceWtIncreaseGenericBlock} as expanded on in the next point.
    \item \textbf{Deviations from ideality $q_m$:} Given a choice of ideal POVM $\finPOVM[\ideal]$, the optimal choice of $q_m$ is simply the minimum value that still satisfies \cref{eq:deviationDefBlock}. This can be seen by noting that the weight in the flag space consists entirely of classical states. Thus, for most quantum information applications, it is desirable to maximise the weight in the preserved subspace. This is a monotonic function in the deviation $q_m$ as seen from \cref{eq:subspaceWtIncreaseGenericBlock}.
\end{enumerate}


\begin{technical}
    If $\infPOVM[\ideal]$ already contains a flag space, then the extra flag space added in \cref{eq:addedFlags} is unnecessary. Instead, the existing flag space can be used to accommodate the imperfections, as seen in \cref{thm:lossNoiseChannel}. 
\end{technical}

\section{Application to QKD} \label{sec:appToQKD}

In this section, we demonstrate the utility of our noise channel framework by applying it to the security analysis of QKD protocols with imperfect threshold detectors. We summarise the impact of our work across security proof techniques in \cref{tab:imperfect-threshold-setups}. Our work significantly advances practical security proofs based on the EAT \cite{dupuis_entropy_2020,metger_generalised_2022,arqand2025marginal}. In particular, our work allows for the treatment of arbitrary, imperfect passive linear optical detection setups via the use of \cref{thm:FSSFineGrained,thm:lossNoiseChannel} with the flag-state squasher \cite{zhang_security_2021}. This usage of the flag-state squasher requires a bound on the weight in the flag space for an imperfect detection setup, as given in \cite[Section III]{wang2025phase}, which extends the work in \cite[Section VII. F.]{kamin2024improved} to the case with loss. Additionally, we demonstrate the application of \cref{thm:genericNoiseChannel} to active basis choice BB84 in \cref{subsec:EATActiveBB84}. \cref{thm:genericNoiseChannel} can be applied without the flag-state squasher and enables the application of the simple squasher \cite{gittsovich_squashing_2014} after the noise channel. This is a large improvement over past security analyses \cite{kamin2025r} which need to assume no dark counts and identical detection efficiencies.

Alternate security proofs using phase error estimation with imperfect active \cite{tupkary2024phaseerrorrateestimation} and passive \cite{wang2025phase} detection setups already exist. However, as can be seen in \cite[Chapter 6]{nahar_phd_2025}, the use of noise channels improves on the key rates obtained for active detection setups. We expect to see similar improvements for passive detection setups as well.

In \cref{sec:appToPS}, we explain how to apply our results to the postselection technique \cite{christandl2009postselection, nahar_postselection_2024}.
In particular, the postselection technique is not directly compatible with the flag-state squasher due to the ad hoc constraint described in \cref{eq:wtOutsideSubspaceGeneric}. However, as we elaborate in \cref{sec:appToPS}, it is possible to indirectly apply \cref{thm:FSSFineGrained,thm:lossNoiseChannel} --- together with the weight-preserving flag-state squasher (WPFSS) \cite[Lemma 7]{nahar_postselection_2024} --- to prove security via the postselection technique. However, this cannot be used with active detection setups as computing the bound on the weight in the flag space for active detection setups is still an open problem.

\begin{table}[t]
  \centering
  \small
  \setlength{\tabcolsep}{4pt}        
  \renewcommand{\arraystretch}{1.35} 
  \begin{tabular}{p{0.15\textwidth}p{0.21\textwidth}p{0.21\textwidth}p{0.33\textwidth}}
    \toprule
    \textbf{Proof technique} &
    \textbf{Earlier work: active basis choice} &
    \textbf{Earlier work: passive basis choice} &
    \textbf{With noise channels} \\
    \midrule

    \textbf{EAT \cite{arqand2025marginal,dupuis_entropy_2020,metger_generalised_2022}} &
    Assumed detection setups with no dark counts and identical detection efficiencies allowing the use of the simple squasher. &
    Assumed exactly known POVM elements, with small number of modes. Used flag-state squasher. &
    Assumes detector parameters are within known ranges, and multi-mode model as stated in \cref{subsec:multimode}. 
    \begin{itemize}[nosep,leftmargin=*]
        \item \cref{thm:genericNoiseChannel} can be used with simple squasher for active detection setups (see \cref{subsec:EATActiveBB84} for an example).
        \item \cref{thm:FSSFineGrained,thm:lossNoiseChannel} can be used with the flag-state squasher for passive detection setups.
    \end{itemize}
    \\
    \midrule
    
    \textbf{Postselection technique \cite{christandl2009postselection,nahar_postselection_2024}} &
    Assumed detection setups with no dark counts and identical detection efficiencies allowing the use of the simple squasher. &
    Assumed exactly known POVM elements, with small number of modes. Used (weight-preserving) flag-state squasher. &
    Assumes detector parameters are within known ranges, and small number of modes.
    \begin{itemize}[nosep,leftmargin=*]
        \item No improvement for active detection setups (see \cref{sec:appToPS}).
        \item \cref{thm:FSSFineGrained,thm:lossNoiseChannel} can be used with the flag-state squasher for passive detection setups (see \cref{sec:appToPS}).
    \end{itemize}
    \\
    \midrule

    \textbf{Phase error-estimation \cite{tomamichel_largely_2017,koashi2009simple}} &
    Assumed detector parameters are within known ranges, and multi-mode model as stated in \cref{subsec:multimode} \cite{tupkary2024phaseerrorrateestimation}.  &
    Assumes detector parameters are within known ranges, and multi-mode model as stated in \cref{subsec:multimode} \cite{wang2025phase}.  &
    Assumes detector parameters are within known ranges, and multi-mode model as stated in \cref{subsec:multimode}.
    \begin{itemize}[nosep,leftmargin=*]
        \item Improved key rates for active detection setups \cite[Chapter 6]{nahar_phd_2025}.
        \item Similar improvements expected for passive detection setups.
    \end{itemize}
    \\

    \bottomrule
  \end{tabular}
  \caption{The impact of noise channels on QKD security proofs for threshold detection setups across proof techniques.}
  \label{tab:imperfect-threshold-setups}
\end{table}

\subsection{EAT for active basis-choice BB84} \label{subsec:EATActiveBB84}

We now describe the detailed application of our methods to the active basis-choice BB84 protocol using the EAT-based proof technique.
Prior to our work, numerical proof techniques such as those based on EAT could not rigorously analyse this protocol in the presence of detector imperfections, such as basis efficiency mismatch and non-zero dark count rates. This is because there is no known way to bound the weight outside the preserved subspace for active basis-choice BB84\footnote{There is some work \cite{trushechkinSecurity2022} to bound this weight. However, this work is in the asymptotic regime, and the form of the bound does not directly match the form used in the numerical proofs \cite{kamin2025finite,kamin2025r}.}, so the flag-state squasher could not be rigorously applied. Meanwhile the simple squasher could not be applied in the presence of basis efficiency mismatch (and non-zero dark counts).

Our method overcomes this limitation for cases where detector efficiencies and dark count rates lie within known, bounded ranges. The process proceeds in three steps. First, common channel loss is attributed to Eve \cite[Fig. 1]{koashi2006efficient}, which, without loss of generality, allows one detector to be assumed to have unit efficiency.
Second, the noise channel constructed in \cref{thm:genericNoiseChannel} is applied. This reduces the security analysis to one with a detection setup with no loss and no dark counts, but where the POVM elements also have flags.
Finally, the component in the unflagged subspace with no loss and no dark counts is further reduced by applying the simple squasher \cite[Theorem 10]{gittsovich_squashing_2014} to obtain a POVM that only acts on a qubit space with loss (and flags). This process is depicted in \cref{fig:EATActive}.
\begin{figure*}
    \centering
    \scalebox{0.8}{\begin{tikzpicture}

\begin{scope}[shift = {(1,0)}]
    \pic {detector={detinf, {$\infPOVM[\dB, \loss]$}, black}};
    
    \draw (-5,0) -- ([xshift=-0.05cm]detinf.west); 
\end{scope}

    \begin{scope}[shift={(-1,-4)}]
        \node[process,xshift = -1cm](NoiseChannel){$\noiseChannel{q(\dB,\loss)}$};
        \pic [right of = NoiseChannel,xshift = 2cm]{detector={detfinideal, $\finPOVM$\\$W^{\dB, \loss}$,black}};
        
        \draw (-3,0) -- (NoiseChannel);
        \draw (NoiseChannel) -- (detfinideal.west);
    \end{scope}

    \begin{scope}[shift={(-1,-8)}]
        \node[process,xshift = -1cm](NoiseChannel){$\noiseChannel{q(\dB,\loss)}$};
        \node[process,right of = NoiseChannel,xshift = 2cm](SqMap){$\Lambda$};
        \pic [right of = SqMap,xshift = 2cm]{detector={detfinideal, $\finPOVM[Q]$\\$W^{\dB, \loss}$,black}};
        
        \draw (-3,0) -- (NoiseChannel.west);
        \draw (SqMap) -- (NoiseChannel);
        \draw (SqMap) -- (detfinideal.west);
    \end{scope}
    
    \node at (0,-2) {\scalebox{2}{$\Big\Updownarrow$}};
    \node at (-0.5,-2) {(i)};
    \node[text=red] at (0.5,-2) {\scalebox{2}?};
    
    \node at (0,-6) {\scalebox{2}{$\Big\Updownarrow$}};
    \node at (-0.5,-6) {(ii)};
    
\end{tikzpicture}}
    \caption{A set of equivalences (as quantum-to-classical measurement channels) of the active basis-choice BB84 detection setups to consider dark counts and loss as part of the noise channel, and to reduce the analysis to a finite-dimensional POVM.} \label{fig:EATActive}
\end{figure*}

Note that the real POVM $\infPOVM[\dB,\loss]$, fixed by the detection setup, is block-diagonal in photon number. In particular, we consider two blocks,
\begin{equation*}
    \Gamma^{\dB,\loss}_{i} = \Gamma^{\dB,\loss}_{i,\,0} \oplus \Gamma^{\dB,\loss}_{i,\,>0}.
\end{equation*}
Similarly, the POVM $\finPOVM$ after the application of the noise channel is fixed by equivalence (ii), and is block-diagonal in photon number (and flags) as
\begin{equation*}
    F_i = \Gamma_{i,\,0}\oplus \Gamma_{i,\,>0}\oplus \ketbra{i}.
\end{equation*}
The application of the noise channel described in \cref{thm:genericNoiseChannel} thus requires us to compute deviations from ideality $q_0$ and $q_{>0}$ such that
\begin{equation}
    \begin{aligned}
        \Gamma^{\dB,\loss}_{i,\,0} &\geq (1-q_0) \Gamma_{i,\,0}\\
        \Gamma^{\dB,\loss}_{i,\,>0} &\geq (1-q_{>0}) \Gamma_{i,\,>0}
    \end{aligned} \quad \forall \quad i.
\end{equation}

As shown in \cref{appsec:noiseChannelDeviationPhaseError}, we can compute these deviations as 
\begin{equation} \label{eq:deviationBoundsActive}
    \begin{aligned}
        q_0 &= 1-(1-\dmax)^2\\
        q_{>0} &= 1-\etamin\left(1-\frac{\dmax}{2}\right),
    \end{aligned}
\end{equation}
where $\dmax$ is the maximum possible dark count rate, and $\etamin$ is the minimum allowed value of the detector efficiency across both detectors (after pulling out common loss). As we typically work in the regime where $\etamin \leq 1-\dmax$, $\qbasis\geq q_0$, the weight in the flag subspace $\H_F$ is given by\footnote{If instead $q_0 \geq \qbasis$, a very similar analysis can still be performed where we must upper bound the weight in the $0$-photon subspace instead. In our work, we use the trivial upper bound of 1 for this case, i.e. $\noiseChannel{q(\dB,\loss)}^\dag\left[\Pi_F\right] \leq q_0 \mathbb{I}$.}
\begin{equation*}
    \begin{aligned}
        \noiseChannel{q(\dB,\loss)}^\dag\left[\Pi_F\right] &\leq q_0 \Pi_0 + \qbasis\Pi_{>0}\\
        &= q_0 + (\qbasis-q_0)\Pi_{>0},
    \end{aligned}
\end{equation*}
for any input state $\rho$.
This requires an upper bound on the weight outside the $0$-photon subspaces. We show in \cref{appsec:minEigenvaluePhaseError} that
\begin{equation*}
    \begin{aligned}
        \Pi_{>0} &\leq \frac{1}{\etamin} \left(\mathbb{I}-\Gamma_{\nc}^{\dB,\loss}\right),
    \end{aligned}
\end{equation*}
where $\Gamma_{\nc}$ is the POVM element corresponding to no detector clicking.
Thus, the weight in the flag subspace after the application of the noise channel can be bounded as
\begin{equation} \label{eq:flagWtActiveBB84}
    \begin{aligned}
        \noiseChannel{q(\dB,\loss)}^\dag\left[\Pi_F\right] &\leq q_0\mathbb{I} + \frac{(\qbasis-q_0)}{\etamin} \left(\mathbb{I}-\Gamma_{\nc}^{\dB,\loss}\right)\\
        &= \left(1-(1-\dmax)^2\right)\mathbb{I} + \left(\frac{(1-\dmax)^2}{\etamin}-\left(1-\frac{\dmax}{2}\right)\right) \left(\mathbb{I}-\Gamma_{\nc}^{\dB,\loss}\right).
    \end{aligned}
\end{equation}

Thus, the numerical security proof can be applied with the finite-dimensional POVM $\finPOVM$ with no dependence on the dark count rates or detector efficiencies; where the weight in the flag-space can be bounded by \cref{eq:flagWtActiveBB84}.


\subsection{Application to the postselection technique} \label{sec:appToPS}

We will first explain the problem of applying \cref{thm:FSSFineGrained,thm:lossNoiseChannel,thm:genericNoiseChannel} to the postselection technique at an intuitive level before formally stating the problem and solution.

The postselection technique is a security proof technique that follows a route that reduces a a security proof against the most general attack to a security proof against a limited IID attack. This reduction incurs a dimensional-dependent penalty to both the security parameter, as well as the key length. In order to facilitate its application to optical protocols, a squashing map must be used \cite[Lemma 6]{nahar_postselection_2024} to reduce the dimension of the problem, and hence reduce the penalty. However, the flag-state squasher uses the subspace estimation (\cref{eq:wtOutsideSubspaceGeneric}) from the infinite-dimensional state\footnote{Note that there exists a de Finetti theorem for infinite-dimensional states \cite{renner2009finetti} that can be used with the subspace estimation, although we expect it to perform worse than the postselection theorem due to the `sacrifice' bits (see figure 2 from Ref.~\cite{sheridan2010finite} for an example comparison between the performance of the de Finetti theorem and postselection technique).}.
Thus, as explained in \cite[Section IV.B.1]{nahar_postselection_2024}, in order to use the flag-state squasher with the postselection technique, one can follow the route first to use the weight-preserving flag-state squasher (WPFSS)\footnote{For our application here, it is not critical to understand all the details of the WPFSS. The only important detail is that the WPFSS does not allow for the usage of any fine-grained click-pattern that constitutes the event $e$, instead bundling them up into a single `coarse-grained' POVM element.} to facilitate the use of the postselection technique, and then the flag-state squasher can be used on the resulting POVM. Note the two distinct uses of squashing maps --- first, the WPFSS \emph{before} using the postselection technique, and then the flag-state squasher (acts on the POVM already squashed by the WPFSS) \emph{after} using the postselection technique.

\begin{technical}
    The WPFSS or simple squasher \cite{gittsovich_squashing_2014} \emph{must} be used to prove security with the postselection technique. Thus, within the postselection technique, the security for active basis-choice protocols cannot be proved in a similar manner to the analysis in \cref{subsec:EATActiveBB84}. In particular, \cref{thm:genericNoiseChannel} cannot be applied before the WPFSS or simple squasher \cite{gittsovich_squashing_2014}, as was necessary in \cref{subsec:EATActiveBB84}.
\end{technical}

Now, it might be expected that \cref{thm:FSSFineGrained,thm:lossNoiseChannel} can be used after the flag-state squasher (which was used on the POVM squashed by the WPFSS) to prove security against untrusted loss and dark counts.
However, the WPFSS requires that all the `fine-grained' click patterns that constitute the `coarse-grained' event $e$ used to find the bound in \cref{eq:wtOutsideSubspaceGeneric} be written as a single POVM element before it can be used. As a result, the flag-state squasher is not applied on the full `fine-grained' POVM consisting of all click patterns; it is instead applied on a `coarse-grained' POVM that bundles together some set of click-patterns into a single element.
Thus, we need to carefully check that the dark count post-processing on the coarse-grained POVM satisfies \cref{eq:11morethan00,eq:noWhiteCounts,eq:sIsS} before using \cref{thm:FSSFineGrained} to this situation.

We will first formally explain what it means to `coarse-grain' the POVM so that we can formally state the proof idea.
A `coarse-graining' is a classical post-processing step whose corresponding stochastic matrix only consists of $1$s and $0$s. For example, consider the passive BB84 detection setup and classically post-process all click patterns that consist of more than a single click into a single `multi-click' event. The stochastic matrix corresponding to this post-processing is given by
\begin{align} \label{eq:cgForBB84}
    \Mcg = \begin{pNiceMatrix}
                    1 & 0 & 0 & 0 & 0 & \Cdots & 0 \\
                    0 & 1 & 0 & 0 & 0 & \Cdots & 0 \\
                    0 & 0 & 1 & 0 & 0 & \Cdots & 0 \\
                    0 & 0 & 0 & 1 & 0 & \Cdots & 0 \\
                    0 & 0 & 0 & 0 & 1 & \Cdots & 1 
                    \CodeAfter
                    \OverBrace[shorten,yshift=0pt]{1-5}{5-7}{12}
                \end{pNiceMatrix}.
\end{align}

The general framework is shown in \cref{fig:equivalencesWPFSS}. It is similar to the framework described in \cref{sec:proofTechnique}.
\begin{figure*}
    \centering
    \scalebox{0.8}{\begin{tikzpicture}
    \pic at (3,0) {detector={detinf, {$\infPOVM[\dB, \loss]$}, black}};

    \node[process, olive, right of = detinf, xshift = 3cm] (PP){$\Mcg'$};
    
    \draw (-5,0) -- ([xshift=-0.05cm]detinf.west); 

    \draw[dashed] ([xshift=0.22cm]detinf.east) -- (PP);
    
    \begin{scope}[shift={(2,-4)}]
        \pic {detector={detinfideal, $\infPOVM[\loss]$,black}};
        \node[process, orange, right of = detinfideal, xshift = 1.5cm](DCPP){$\PP^{\dB}$};
        \node[process, olive, right of = DCPP, xshift = 1.5cm] (SqPP){$\Mcg'$};
        
        \draw (-7,0) -- ([xshift=-0.25cm]detinfideal.west);
        \draw[dashed] ([xshift=0.35cm]detinfideal.east) -- (DCPP);
        \draw[dashed] (DCPP.east) -- (SqPP);
    \end{scope}

    \begin{scope}[shift={(2,-8)}]
        \pic {detector={detinfideal2, $\infPOVM[\loss]$,black}};
        \node[process, orange, right of = detinfideal2, xshift = 1.5cm](SqPPNew){$\Mcg$};
        \node[process, olive, right of = SqPPNew, xshift = 1.5cm] (DCPPNew){$\Mdc$};
        
        \draw (-7,0) -- ([xshift=-0.25cm]detinfideal2.west);
        \draw[dashed] ([xshift=0.35cm]detinfideal2.east) -- (SqPPNew);
        \draw[dashed] (SqPPNew.east) -- (DCPPNew);
    \end{scope}

    \begin{scope}[shift={(2,-11.5)}]
        \node[process](SqMap){$\Lambda_{\text{WPFSS}}$};
        \pic [right of = SqMap,xshift = 1cm]{detector={detfinideal, $\finPOVM[\loss]$,black}};
        \node[process, olive, right of = detfinideal, xshift = 2cm] (DCPPNew){$\Mdc$};
        
        \draw (-7,0) -- (SqMap.west);
        \draw (SqMap) -- ([xshift=-0.25cm]detfinideal.west);
        \draw[dashed] ([xshift=0.35cm]detfinideal.east) -- (DCPPNew);
    \end{scope}

    \begin{scope}[shift={(-1,-15)}]
        \node[process](WPFSS){$\Lambda_{\text{WPFSS}}$};
        \node[process,right of = WPFSS,xshift = 2cm](FSS){$\Lambda_{\text{FSS}}$};
        \pic [right of = FSS,xshift = 1cm]{detector={detfinideal, $\finPOVM[\loss]$\\$W$,black}};
        \node[process, olive, right of = detfinideal, xshift = 2cm] (DCPPNew){$\Mdc$};
        
        \draw (-4,0) -- (WPFSS.west);
        \draw (WPFSS.east) -- (FSS.west);
        \draw (FSS) -- ([xshift=-0.25cm]detfinideal.west);
        \draw[dashed] ([xshift=0.35cm]detfinideal.east) -- (DCPPNew);
    \end{scope}

    \begin{scope}[shift={(-2,-18.5)}]
        \node[process](WPFSS){$\Lambda_{\text{WPFSS}}$};
        \node[process,right of = WPFSS,xshift = 2cm](FSS){$\Lambda_{\text{FSS}}$};
        \node[process,right of = FSS,xshift = 2cm](NoiseChannel){$\Phi_{\dB}$};
        \pic [right of = NoiseChannel,xshift = 2cm]{detector={detfinideal, $\finPOVM[\loss]$\\$W^{\dB}$,black}};
        
        \draw (-3,0) -- (WPFSS.west);
        \draw (WPFSS.east) -- (FSS.west);
        \draw (FSS) -- (NoiseChannel);
        \draw (NoiseChannel) -- ([xshift=0cm]detfinideal.west);
    \end{scope}

    \begin{scope}[shift={(-2,-22)}]
        \node[process,xshift = -1cm](WPFSS){$\Lambda_{\text{WPFSS}}$};
        \node[process,right of = WPFSS,xshift = 1.5cm](FSS){$\Lambda_{\text{FSS}}$};
        \node[process,right of = FSS,xshift = 1.5cm](DCNoiseChannel){$\Phi_{\dB}$};
        \node[process,right of = DCNoiseChannel,xshift = 1.5cm](LossNoiseChannel){$\Phi_{\loss}$};
        \pic [right of = LossNoiseChannel,xshift = 1.5cm]{detector={detfinideal, $\finPOVM[\loss]$\\$W^{\dB,\loss}$,black}};
        
        \draw (-3,0) -- (WPFSS.west);
        \draw (WPFSS.east) -- (FSS.west);
        \draw (FSS) -- (DCNoiseChannel);
        \draw (DCNoiseChannel) -- (LossNoiseChannel);
        \draw (LossNoiseChannel) -- ([xshift=0cm]detfinideal.west);
    \end{scope}
    
    \node at (3,-2) {\scalebox{2}{$\Big\Updownarrow$}};
    \node at (2.5,-2) {(i)};
    
    \node at (3,-6) {\scalebox{2}{$\Big\Updownarrow$}};
    \node at (2.5,-6) {(ii)};
    \node[text=red] at (3.5,-6) {\scalebox{2}?};

    \node at (3,-10) {\scalebox{2}{$\Big\Updownarrow$}};
    \node at (2.5,-10) {(iii)};

    \node at (3,-13.4) {\scalebox{2}{$\Big\Updownarrow$}};
    \node at (2.5,-13.4) {(iv)};

    \node at (3,-17) {\scalebox{2}{$\Big\Updownarrow$}};
    \node at (2.5,-17) {(v)};
    \node[text=red] at (3.5,-17) {\scalebox{2}?};

    \node at (3,-20.2) {\scalebox{2}{$\Big\Updownarrow$}};
    \node at (2.5,-20.2) {(vi)};
\end{tikzpicture}}
    \caption{A set of equivalences (as quantum-to-classical measurement channels) of detection setups to use \cref{thm:FSSFineGrained,thm:lossNoiseChannel} with the postselection technique. Here, $\Mcg'$ represents the classical post-processing carried out in the protocol, $\PPdb$ is the dark count post-processing, $\Mcg$ is the coarse-graining required for the WPFSS $\Lambda_{\text{WPFSS}}$, and $\Mdc$ is the post-processing fixed by \cref{eq:WPFSSPPSwap}. In the figure, olive post-processing maps represent free choices that can be made to make equivalences (ii) and (v) hold. Orange post-processing maps represent stochastic processes fixed by equivalences (i) and (iii).} \label{fig:equivalencesWPFSS}
\end{figure*}
Once again, each equivalence is an equivalence of quantum-to-classical measurement channels.
\begin{itemize}
    \item We start by considering a generic detection setup that uses threshold detectors with dark counts and loss. Additionally, we allow for some classical post-processing $\Mcg'$. Although, this can be chosen to be any arbitrary post-processing, we choose it to be a coarse-graining for simplicity.
    \item Equivalence (i) follows from the fact that dark counts can be modelled as a classical post-processing $\PPdb$ as described in \Cref{subsec:DCPP}.
    \item Equivalence (ii) must be shown to hold, similar to the discussion in \cref{subsubsec:swapPPCond}; we need to find some stochastic matrix $\Mdc$ such that
    \begin{align} \label{eq:WPFSSPPSwap}
        \Mcg' \PPdb = \Mdc \Mcg.
    \end{align}
    Here $\Mcg$ is the coarse-graining required for the WPFSS.
    \item Equivalence (iii) follows from the existence of the WPFSS \cite[Lemma 7]{nahar_postselection_2024}. The postselection technique can then be used on the resulting finite-dimensional systems.
    \item Equivalence (iv) follows from the existence of the flag-state squasher \cite[Theorem 1]{zhang_security_2021}.
    \item Equivalence (v) follows from \cref{thm:FSSFineGrained}, if $\Mdc$ satisfies \cref{eq:sIsS,eq:noWhiteCounts,eq:11morethan00}.
    \item Equivalence (vi) follows from \cref{thm:lossNoiseChannel}, assuming that the coarse-graining $\Mcg$ used in the WPFSS preserves the single-click and no-click events, as is commonly the case.
\end{itemize}
The validity of equivalences (ii) and (v) depends on the coarse-graining $\Mcg$ needed for the WPFSS.
As explained above, the coarse-graining $\Mcg$ must have the event used to bound the weight outside the preserved subspace (\cref{eq:wtOutsideSubspaceGeneric}) as one of the POVM elements. In general, for different protocol-dependent choices \cite{li_improving_2020,nahar_imperfect_2023} of the weight estimation event, these equivalences would have to be checked on a case-by-case basis. Fortunately, for the more generic choice of weight estimation event described in \cite{kamin2024improved,wang2025phase}, we can check the validity of these equivalences \emph{independent} of protocol choice, which we now proceed to discuss.

The weight estimation in \cite{kamin2024improved,wang2025phase}, which can be applied to any passive detection setup, uses the event consisting of any click pattern with more than a single-click.
The coarse-graining $\Mcg$ corresponding to this weight estimation event does nothing to the no-click and single-click event, but considers the rest of the events to be a single multi-click event. Concretely, the stochastic matrix corresponding to this coarse-graining is given by
\begin{align} \label{eq:cgGeneric}
    \Mcg = \begin{pNiceArray}{c|ccc|ccc}
                1 & 0 & \Cdots & 0 & 0 & \Cdots & 0 \\
                \hline
                0 & \Block{3-3} <\Large>{\mathbb{I}_\nS} & & & \Block{3-3}<\large>{\mathbf{0}_{\nS \times \nM}}\\
                \Vdots & & & & \\
                0 & & & & \\
                \hline
                0 &  0 & \Cdots & 0 &  1 & \Cdots & 1
                \CodeAfter
                \UnderBrace[shorten,yshift=-0pt]{1-5}{5-7}{\nM}
                \UnderBrace[shorten,yshift=-0pt]{1-2}{5-4}{\nS}
            \end{pNiceArray},
\end{align}
\newline
\noindent where $\nS$ is the number of single-click events, and $\nM$ is the number of multi-click events.

For this coarse-graining, we shall verify that there exists a classical post-processing $\Mdc$ that satisfies \cref{eq:WPFSSPPSwap}.
First, we set the coarse-graining $\Mcg'$ performed in the protocol to be the same as the coarse-graining $\Mcg$ described in \cref{eq:cgGeneric}.
Further, we need to make the physically motivated assumption that the dark count post-processing $\PPdb$ does not turn multi-clicks into no-click or single-click events, i.e.
\begin{align}
    \PPdb{i}{m} = 0,
\end{align}
for all $i \in \{0\} \cup \singleSet$, and $m \in \multiSet$. Thus, the dark count post-processing can be written as a block-matrix whose top-right block is zero as

\begin{align} \label{eq:dcPPGeneric}
    \PPdb = \begin{pNiceArray}{c|c}
                {\PPdb{\multiSet^C}{\multiSet^C}} & {\mathbf{0}_{(\nS+1) \times \nM}}\\
                \hline
                {\PPdb{\multiSet}{\multiSet^C}} &{\PPdb{\multiSet}{\multiSet}}
                \CodeAfter
                \OverBrace[shorten,yshift=-0pt]{1-1}{2-1}{\nS+1}
                \OverBrace[shorten,yshift=-0pt]{1-2}{2-2}{\nM}
            \end{pNiceArray}.
\end{align}
Importantly, the $\mathbf{0}_{(\nS+1) \times \nM}$ block along with the fact that $\PPdb{\multiSet}{\multiSet}$ is a stochastic matrix can be used to verify from explicit computation that the following ansatz satisfies \cref{eq:WPFSSPPSwap}:

\begin{align}
    \Mdc = \begin{pNiceArray}{ccc|c}
                \Block{3-3}<\Large>{\PPdb{\multiSet^C}{\multiSet^C}} & & & 0\\
                & & & \Vdots\\
                & & & 0\\
                \hline
                s^1_{\multiSet \vert \multiSet^C} & \Cdots & s^{\nS+1}_{\multiSet \vert \multiSet^C} & {1}
                \CodeAfter
                \OverBrace[shorten,yshift=-0pt]{1-1}{3-3}{\nS+1}
            \end{pNiceArray},
\end{align}
where $s^i_{\multiSet \vert \multiSet^C}$ is the sum of all elements in the $i^\text{th}$ column of $\PPdb{\multiSet}{\multiSet^C}$.

Moreover, it is straightforward to verify that this post-processing $\Mdc$ satisfies \cref{eq:sIsS,eq:noWhiteCounts,eq:11morethan00}. Thus, \cref{thm:FSSFineGrained} can be used to show that equivalence (v) holds.
We have hence shown that the postselection technique can be used with the results of this paper with an appropriate choice of coarse-graining $\Mcg$.

\begin{technical}[Application to dimension reduction method]
    The dimension reduction method \cite{upadhyaya_dimension_2021} is a generic tool that bounds the optimal value of an SDP $A$ with a smaller SDP $A_\text{proj}$ constructed by projecting the optimsation variables used in SDP $A$ into a smaller subspace. Thus, it is a useful tool to reduce the computational resources used for numerical key rate computations.
    
    As it is a simplification at the level of the single-round key rate semidefinite program (SDP), it can be directly used with the results of this paper as an additional component after constructing the noise channel with the help of the flag-state squasher.
This simplifies the numerical computations significantly \cite[Fig. 9]{upadhyaya_dimension_2021}.
\end{technical}

\subsection{Illustrative example application: Three state time-bin encoded protocol} \label{subsec:3stateExample}

We now illustrate the impact of addressing detector imperfections via \cref{thm:FSSFineGrained,thm:genericNoiseChannel} by plotting key rates for the time-bin encoded three-state protocol. We will use the postselection technique as described in \cite{nahar_postselection_2024,nahar_phd_2025}.

\subsubsection{Optical setup} \label{sec:ThreeStateOpticalSetup}

\begin{figure*}[t]
    \centering
    \includegraphics[width = \linewidth]{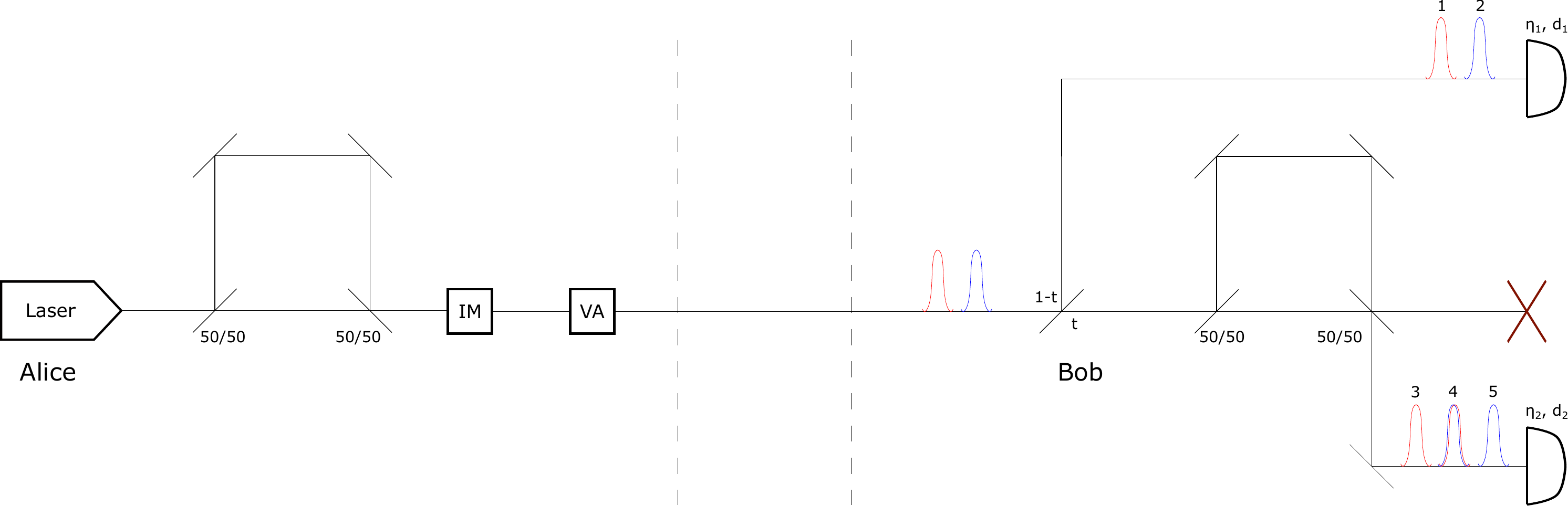}
    \caption{Schematics of the implementation of the three-state protocol as in \cite{boaron_secure_2018}. The numbers below the beam splitters reflect their transmissivity. IM and VA refer to intensity modulator and variable attenuator respectively.} \label{fig:3StateExpSetup}
\end{figure*}

We first describe the optical setup of the time-bin encoded three-state protocol depicted in \cref{fig:3StateExpSetup}.

We consider a protocol where Alice uses a laser source to prepare a phase-randomised coherent state in two time-bin modes as follows:
\begin{itemize}
    \item[] $0$, $\mu$: $\ket{0}\otimes\ket{\sqrt{\mu} e^{i\theta}}$
    \item[] $1$, $\mu$: $\ket{\sqrt{\mu} e^{i\theta}}\otimes\ket{0}$
    \item[] $+$, $\mu$: $\ket{\frac{\sqrt{\mu} e^{i\theta}}{\sqrt{2}}}\otimes\ket{\frac{\sqrt{\mu} e^{i\theta}}{\sqrt{2}}}$,
\end{itemize}
where the phase $\theta\in[0,2\pi)$ is uniformly random. The variable attenuator and intensity modulator serve to choose the values of the encoding $\{0,1,+\}$ and intensity $\mu$.
Bob's measurement setup is identical to that considered in \cite{nahar_imperfect_2023, boaron_secure_2018} as depicted in \cref{fig:3StateExpSetup}. This consists of a passive beam splitter which chooses between the `Z-basis' arm, which measures the time-of-arrival of the pulses, and the `X-basis' arm, which consists of a Mach-Zehnder interferometer. Note that the Mach-Zehnder interferometer only uses one detector to minimise implementation costs.

The basis choice beam splitter has reflectivity $t\in[\tmin,\tmax]$. Both detectors here are threshold detectors, with efficiencies $\eta_1,\eta_2\in[\etamin,\etamax]$ and dark count rates $d_1,d_2\in[0,\dmax]$. The exact values of the ranges are parameters that will be set later. For simplicity, we assume that the 50:50 beam splitters that form a part of the Mach-Zehnder interferometer are perfect, though variations of the same ideas we use for the rest of the imperfections would extend to these imperfections as well.

\subsubsection{Applying the WPFSS and flag-state squasher} \label{subsec:WPFSSThreeState}

To apply the WPFSS, we define the weight-estimation event $e$ to be any click pattern that records a click in both detectors, while ignoring all clicks in the middle time slot of the Mach-Zehnder interferometer. That is, the event corresponding to click patterns which contain at least one of time bin $1$ or $2$, and contain at least one of $3$ or $5$ (see \cref{fig:3StateExpSetup}). Note that this choice is different from the more generic choice considered in \cref{sec:appToPS}, though it matches the choice made in Ref.~\cite{nahar_imperfect_2023}.

Having fewer outcomes leads to a smaller flag space, which is beneficial in reducing the penalty of using the postselection technique (see \cite{nahar_postselection_2024}) and for faster numerics. Thus, we coarse-grain all the outcomes to the outcome in which no detector clicks, the outcomes in which a single detector in a single time bin clicks, the weight-estimation event $e$, and all other outcomes grouped into one. This coarse-graining results in 8 flag states (corresponding to 8 coarse-grained POVM outcomes). For the simulations here, we choose the photon-number cut-off to be $1$.

If we assume unit efficiencies, we can use the results proved in Appendix C of Ref.~\cite{nahar_imperfect_2023}, to find
\begin{align}
    \lambdamin{\overline{\Pi}_{\leq\nFSS} \Gamma_e \overline{\Pi}_{\leq\nFSS}} \geq 1-t^{N+1}-\left(1-\frac{t}{4}\right)^{N+1}+\left(\frac{3t}{4}\right)^{N+1},
\end{align}
where $t$ is Bob's basis-choice beam-splitting ratio.
We now construct an equivalent detection setup with unit efficiencies in order to use a modified version of this result that accounts for the detection setup parameter ranges.

\subsubsubsection{Equivalent detection setup} \label{subsec:detSetupWithoutLoss}

\begin{figure*}[t]
    \centering
    \begin{subfigure}[t]{0.49\linewidth}
        \centering
        \includegraphics[width=\linewidth]{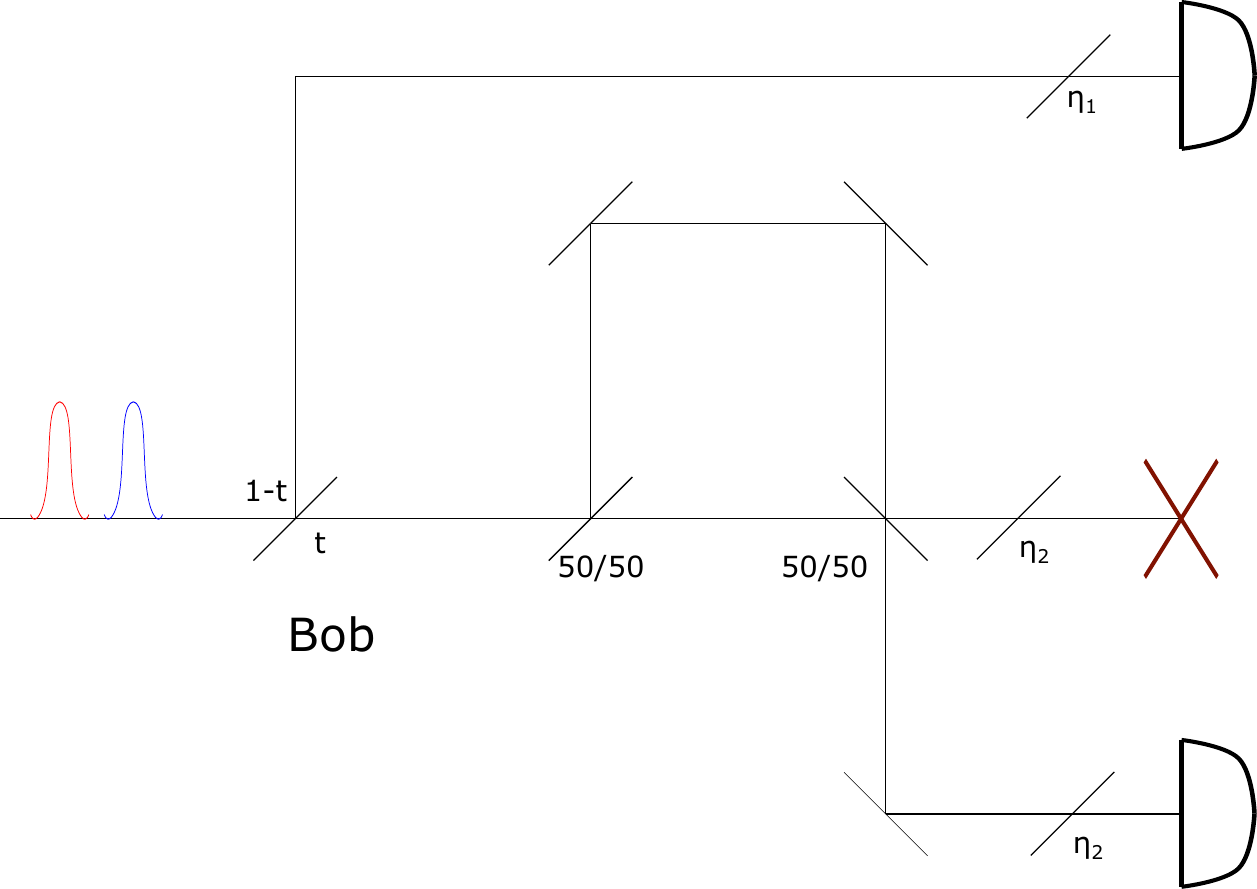}
        \caption{Detectors with efficiency $\eta_i$ represented as beam splitters with transmissivity $\eta_i$ followed by lossless detectors for the three-state protocol detection setup.}
        \label{fig:lossAsBeamSplitters}
    \end{subfigure}%
    \hfill
    \begin{subfigure}[t]{0.49\linewidth}
        \centering
        \includegraphics[width=\linewidth]{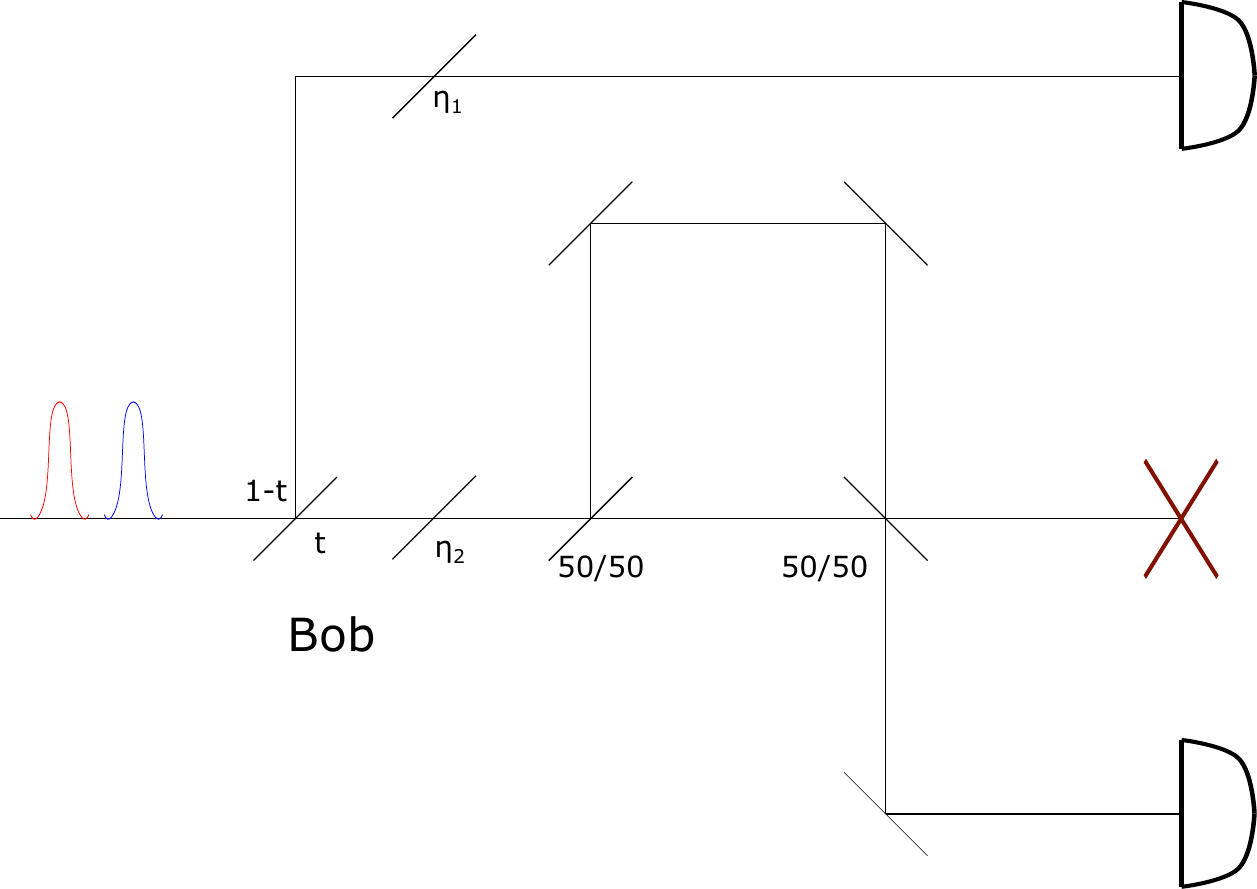}
        \caption{The common loss can be pulled to the input arm of the Mach-Zehnder interferometer.}
        \label{fig:pullCommonMZLoss}
    \end{subfigure}
    \caption{Simplifying the description of the detection setup with lossy detectors.}
    \label{fig:combinedFigure}
\end{figure*}

We first use the characterisation of a detector with efficiency $\eta$ described in \cref{sec:DetectorModel}; that of a beam splitter with transmissivity $\eta$ followed by a lossless detector. The missing detector can always be preceded by a beam splitter of any transmissivity. This is depicted in \cref{fig:lossAsBeamSplitters}.
Since both beam splitters at both outputs of the Mach-Zehnder interferometer have the same transmissivity, this can be equivalently represented as a single beam splitter at the input of the Mach-Zehnder interferometer, as depicted in \cref{fig:pullCommonMZLoss}.

\begin{figure*}[t]
    \centering
    \begin{subfigure}[t]{0.49\linewidth}
        \centering
        \includegraphics[width=\linewidth]{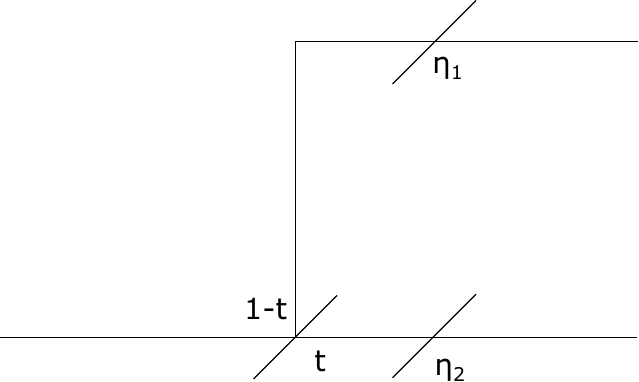}
        \caption{Zooming in on the part of the equivalent detection setup analysed.}
        \label{fig:basisChoicePart}
    \end{subfigure}%
    \hfill
    \begin{subfigure}[t]{0.49\linewidth}
        \centering
        \includegraphics[width=\linewidth]{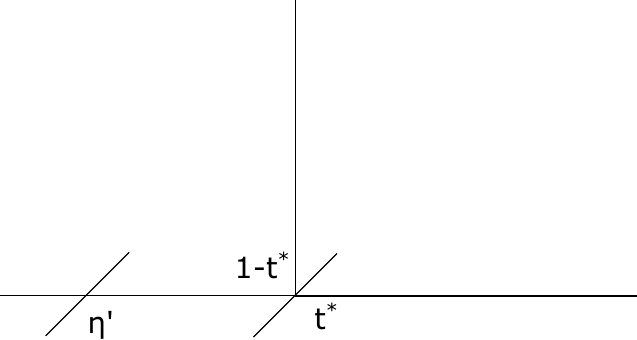}
        \caption{The losses can be pulled to the input arm of the basis-choice beam splitter, with some change in the splitting ratio.}
        \label{fig:pullUncommonLoss}
    \end{subfigure}
    \caption{Simplifying the description of the detection setup with lossy detectors.}
    \label{fig:pullingOutLoss}
\end{figure*}

Next, we use ideas from \cite[Section 1.5]{narasimhachar2011study} to pull both beam splitters that represent the loss before the basis-choice beam splitter. Doing so would also change the basis choice beam-splitting ratio. This is depicted in \cref{fig:pullingOutLoss}.
It is sufficient to show that this equivalence holds for any coherent states, as these form a basis for the entire space (of operators). We establish this as follows. Any coherent state $\ket{\alpha}$ input into the setup depicted in \cref{fig:basisChoicePart} has output $\ket{\sqrt{(1-t)\,\eta_1}\,\alpha}\otimes \ket{\sqrt{t\,\eta_2}\,\alpha}$. Any coherent state $\ket{\alpha}$ input into the setup depicted in \cref{fig:pullUncommonLoss} has output $\ket{\sqrt{\eta'\,(1-\teff)}\,\alpha}\otimes \ket{\sqrt{\eta'\,\teff}\,\alpha}$. This holds if
\begin{align*}
    (1-t)\,\eta_1 &= \eta'\,(1-\teff)\\
    t\,\eta_2 &= \eta'\,\teff.
\end{align*}
Adding both equations gives $\eta' = t\,\eta_2 + (1-t)\,\eta_1$. This is a convex combination of $\eta_1$ and $\eta_2$, and hence lies between the maximum and minimum of the two. Consequently, $\eta'\in[0,1]$ is well-defined. We can use this to solve for the effective beam-splitting ratio $\teff$:
\begin{gather}
    \label{eq:actualTeff} \teff = \frac{t\,\eta_2}{t\,\eta_2 + (1-t)\,\eta_1}\\
    \label{eq:boundOnTeff} \teffmin\coloneqq \tmin\frac{\etamin}{\etamax}\leq \teff \leq \tmax \frac{\etamax}{\etamin}\eqqcolon \teffmax,
\end{gather}
where we have used the ranges of the beam-splitting ratio and efficiencies to bound the effective beam-splitting ratio in \cref{eq:boundOnTeff}. Note that although the upper bound in \cref{eq:boundOnTeff} can be larger than $1$, the true value of the effective beam-splitting ratio $\teff$ given in \cref{eq:actualTeff} cannot be larger than $1$:
\begin{align*}
    \teff &= \frac{t\,\eta_2}{t\,\eta_2 + (1-t)\,\eta_1}\\
    &= \frac{1}{1+\frac{(1-t)\,\eta_1}{t\,\eta_2}}\leq 1.
\end{align*}
Thus, the effective beam-splitting ratio $\teff\in[0,1]$ is also well-defined.

The common loss $\eta'$ can now be considered to be a part of Eve's channel. Thus, we have reduced the detection setup in \cref{fig:3StateExpSetup} to one without loss, but with a modified beam-splitting ratio. Thus, using the results proved in \cite[Appendix C]{nahar_imperfect_2023}, we can obtain the required bound
\begin{align*}
    \lambdamin{\overline{\Pi}_{\leq\nFSS} \Gamma_e \overline{\Pi}_{\leq\nFSS}} &\geq 1-\left(\teff\right)^{N+1}-\left(1-\frac{\teff}{4}\right)^{N+1}+\left(\frac{3\teff}{4}\right)^{N+1}\\
    &\geq 1-\left(\teffmax\right)^{N+1}-\left(1-\frac{\teffmin}{4}\right)^{N+1}+\left(\frac{3\teffmin}{4}\right)^{N+1}.
\end{align*}
As our analysis uses the single-photon cut-off $\nFSS = 1$, we can straightforwardly derive a tighter and simpler bound:
\begin{align}
    \lambdamin{\overline{\Pi}_{\leq1} \Gamma_e \overline{\Pi}_{\leq1}} &\geq 1-\left(\teff\right)^{2}-\left(1-\frac{\teff}{4}\right)^{2}+\left(\frac{3\teff}{4}\right)^{2}\\
    &= \frac{\teff}{2}-\frac{16\left(\teff\right)^{2}+\left(\teff\right)^{2}-9\left(\teff\right)^{2}}{16}\\
    &= \frac{1}{2}\teff\left(1-\teff\right)\\
    \label{eq:lambdaMinBound3State} &\geq \frac{1}{2}\teffmin\left(1-\teffmax\right).
\end{align}
This is the bound that we shall use in our key length computations\footnote{Technically, we also need to worry about the dark counts. However, the dark counts only increase the probability of this click pattern occurring. Thus, this is a valid lower bound with dark counts as well.}.

\subsubsection{Noise channel constructions} \label{subsec:noiseChannelThreeState}

We follow the steps described in \cref{sec:appToPS} to use \cref{thm:FSSFineGrained,thm:genericNoiseChannel} for the detection setup. In particular, we apply the six equivalences in \cref{fig:equivalencesWPFSS}, then address the imperfect beam splitter via \cref{thm:genericNoiseChannel}.
\begin{enumerate}[label={}, leftmargin=0pt, itemindent=0pt, labelindent=0pt, labelsep=0pt, listparindent=0pt]
    \item \textbf{Equivalence (i):} The equivalence follows from the fact that dark counts can be viewed as a post-processing, as described in \cref{sec:DetectorModel}.
    \item \textbf{Equivalence (ii):} Although the coarse-graining used for the weight estimation is slightly different from that considered in \cref{sec:appToPS}, a similar analysis allows us to swap the coarse-graining and the dark count post-processing. The swapped dark count post-processing obtained satisfies \cref{eq:11morethan00,eq:noWhiteCounts,eq:sIsS}, which will be useful in establishing equivalence (v).
    \item \textbf{Equivalence (iii):} The WPFSS can be applied as described in \cref{subsec:WPFSSThreeState}. At this point, the postselection technique \cite{christandl2009postselection,nahar_postselection_2024} can be applied to restrict Eve's attacks to IID.
    \item \textbf{Equivalence (iv):} The flag-state squasher \cite[Theorem 1]{zhang_security_2021} can be applied. We describe the restriction of Eve's attacks in terms of a constraint on the weight $W$ of the IID state shared by Alice and Bob outside the preserved subspace:
    \begin{equation}
        W\coloneqq\Tr\left[\overline{\Pi}_{\leq 1} \rho_{AQ}\right] \leq \frac{\Tr\left[F_e \rho_{AQ}\right]}{\lambdamin{\overline{\Pi}_{\leq1} \Gamma_e \overline{\Pi}_{\leq1}}},
    \end{equation}
    where $\lambdamin{\overline{\Pi}_{\leq1} \Gamma_e \overline{\Pi}_{\leq1}}$ is lower bounded in \cref{eq:lambdaMinBound3State}.
    \item \textbf{Equivalence (v):} This follows from the fact that the swapped dark count post-processing satisfies \cref{eq:11morethan00,eq:noWhiteCounts,eq:sIsS}, as stated in equivalence (ii). Thus, applying \cref{thm:FSSFineGrained} we obtain the weight constraint,
    \begin{equation}
        1-W^{\dB} \coloneqq \Tr\left[\Pi_{\leq 1}\rhodB\right] \geq (1-\dmax)^5\, W,
    \end{equation}
    where $(1-\dmax)^5$ is the minimum probability that none of the $5$ time bins encounter a dark count.
    \item \textbf{Equivalence (vi):} In \cref{subsec:detSetupWithoutLoss}, we constructed an equivalent detection setup which has no loss. Thus, we do not need to use \cref{thm:lossNoiseChannel}. However, the basis-choice beam splitter of this equivalent setup is imperfect, i.e, the effective beam-splitting ratio $\teff$ can take any value in the range $\teff \in [\teffmin,\teffmax]$, as given in \cref{eq:boundOnTeff}. Hence, we apply \cref{thm:genericNoiseChannel} to accommodate this imperfection as described below.
\end{enumerate}
In the preserved subspace $\Pi_{\leq 1}$, all POVM elements other than the single-click and no-click outcomes are given by the $0$ matrix. Thus, the POVM elements, projected onto the preserved subspace $\Pi_{\leq1}$ can be listed as
\begin{equation} \label{eq:threeStatePOVMleq1}
    \begin{aligned}
        \Pi_{\leq1}F_{\nc}^{\teff}\Pi_{\leq1} &= \ketbra{0}\otimes \ketbra{0} + \frac{\teff}{4} \left(\ket{0}\otimes\ket{1}+ \ket{1}\otimes\ket{0}\right)\left(\bra{0}\otimes\bra{1}+ \bra{1}\otimes\bra{0}\right)\\
        \Pi_{\leq1}F_1^{\teff}\Pi_{\leq1} &= \left(1-\teff\right) \ketbra{0}\otimes \ketbra{1}\\
        \Pi_{\leq1}F_2^{\teff}\Pi_{\leq1} &= \left(1-\teff\right) \ketbra{1}\otimes \ketbra{0}\\
        \Pi_{\leq1}F_3^{\teff}\Pi_{\leq1} &= \frac{\teff}{2} \ketbra{0}\otimes \ketbra{1}\\
        \Pi_{\leq1}F_4^{\teff}\Pi_{\leq1} &=  \frac{\teff}{4} \left(-\ket{0}\otimes\ket{1}+ \ket{1}\otimes\ket{0}\right)\left(-\bra{0}\otimes\bra{1}+ \bra{1}\otimes\bra{0}\right)\\
        \Pi_{\leq1}F_5^{\teff}\Pi_{\leq1} &=  \frac{\teff}{2} \ketbra{1}\otimes \ketbra{0}\\
        \Pi_{\leq1}F_{\mathrm{others}}^{\teff}\Pi_{\leq1} &= 0.
    \end{aligned}
\end{equation}
In particular, the POVM elements have the same structure as that of an active basis choice with probability $\teff:(1-\teff)$.
Using
\begin{equation}
    \begin{gathered}
        \frac{\teffmin}{\teffmax}\teffmax = \teffmin \leq \teff\\
        \frac{\teffmin}{\teffmax}(1-\teffmax) \leq \frac{\teffmin}{\teffmax}(1-\teff) \leq (1-\teff),
    \end{gathered}
\end{equation}
In the vacuum subspace, the only non-zero POVM element corresponds to the no-click outcome. As this does not depend on any detection setup parameters, we choose the ideal POVM to be identical to the real POVM in this subspace. Thus, the deviation from ideality $q_0$ in this subspace is $0$.

In the single-photon subspace, it is straightforward to show that $\finPOVM[1;\;\teff]\geq \frac{\teffmin}{\teffmax} \finPOVM[1;\;\teffmax]$. A similar computation results in $\finPOVM[1;\;\teff]\geq \frac{1-\teffmax}{1-\teffmin} \finPOVM[1;\;\teffmin]$.
Thus, we have the freedom to use \cref{thm:genericNoiseChannel} with either ideal POVM $\finPOVM[0]\oplus\finPOVM[1;\;\teffmax]$ or $\finPOVM[0]\oplus\finPOVM[1;\;\teffmin]$, with corresponding deviation metric $q_1 = \frac{\Delta t}{\teffmax}$ or $q_1 = \frac{\Delta t}{1-\teffmin}$, respectively. Here, $\Delta t \coloneqq \teffmax-\teffmin$.

As we consider the case with $\tmax < 0.5$, the choice with the smallest deviation metric $q$ is
\begin{equation}
    \begin{aligned}
        \text{Ideal POVM: } &\finPOVM[0]\oplus\finPOVM[1;\;\teffmin]\\
        \text{Deviation metrics: } &q_0 = 0\\
        &q_1 = \frac{\Delta t}{1-\teffmin}.
    \end{aligned}
\end{equation}

In order to bound the final weight outside the preserved subspace after applying \cref{thm:genericNoiseChannel} we need to upper bound the weight in the single-photon subspace. We do this as follows. Observe from \cref{eq:threeStatePOVMleq1} that $F^{\teff}_Z \coloneqq F^{\teff}_1 + F^{\teff}_2$ is such that $\Pi_1 F^{\teff}_Z \Pi_1 = (1-\teff) \Pi_1$.
Thus,
\begin{align*}
    \Tr\left[\rhodB\Pi_1\right] &= \frac{\Tr\left[\Pi_1 F^{\teff}_Z \Pi_1 \rhodB\right]}{1-\teff}\\
    &\leq \frac{\Tr\left[F^{\teff}_Z \rhodB\right]}{1-\teff}\\
    &\leq \frac{\Tr\left[F^{\teff}_Z \rhodB\right]}{1-\teffmax}
\end{align*}

This results in the final weight constraint,
\begin{align}
    1-W^{\mathrm{final}} &= \Tr\left[\rhodB\Pi_0\right] + (1-q_1)\Tr\left[\rhodB\Pi_1\right]\\
    &= \Tr\left[\rhodB\Pi_{\leq1}\right] - q_1 \Tr\left[\rhodB\Pi_1\right]\\
    &\geq 1- W^\dB - q_1 \frac{\Tr\left[F_Z^{\teff}\rhodB \right]}{1-\teffmax}\\
    &= (1-\dmax)^5\, \left(1-\frac{\Tr\left[F_e \rho_{AQ}\right]}{\frac{1}{2}\teffmin\left(1-\teffmax\right)}\right) - \frac{\Delta t}{(1-\teffmin)(1-\teffmax)}\Tr\left[F_Z \rho_{AQ}\right].
\end{align}

\subsubsection{Results}

We chose the target secrecy parameter $\epsSec = 2 \times 10^{-12}$.
We do not explain the details of the postselection technique here, which is exactly identical to \cite[Chapter 5]{nahar_phd_2025}.
We do not explain the details of the variable-length IID decoy-state security proof here, and instead refer the interested reader to Ref.~\cite{kamin2025improved}. The simulation parameters are listed in \cref{tab:sim-params} for reference. See \cite{nahar_postselection_2024} for a more complete description of the classical steps of the protocol.

\newcommand{\NA}{\multicolumn{1}{|c|}{\rule{0.08\linewidth}{0.4 pt}}}
\begin{table}[htbp!]
    \centering
    \renewcommand{\arraystretch}{1.2}
    \begin{tabular}{|p{4.5cm}|p{3cm}|p{6cm}|}
    \hline
    \textbf{Parameter} & \textbf{Value} & \textbf{Notes} \\
    \hline
    IID secrecy parameter $\epsIID$ &  $\epsIID = {\left(\epsSec/8 g_{n,x}\right)^2}$ & Computed in \cite[Eq. (5.42)]{nahar_phd_2025}. $g_{n,x}$ is the dimensional-dependent penalty associated with the use of the postselection technique. \\
    \hline
    Privacy amplification portion of secrecy parameter $\epsPA$ & $\epsPA = \epsIID/2$ & Quantifies probability of privacy amplification failing. \\
    \hline
    Parameter estimation portion of secrecy parameter $\epsAT$ & $\epsAT = \epsIID/2$ & Quantifies probability of parameter estimation failing. \\
    \hline
    Number of protocol rounds $n$ & $n=10^{12}$ & \NA \\
    \hline
    Test sample size & $0.05\,n$ & Fraction of signals used for testing. \\
    \hline
    Channel attenuation & $0.16~\mathrm{dB/km}$ & Assumed loss-only channel for simplicity. \\
    \hline
    Basis-choice beam-splitting ratio $\bar{t}$ & $\bar{t}=0.1$ & Midpoint of range of possible values. The setup without imperfections uses this as the fixed beam-splitting ratio.\\
    \hline
    Detector efficiencies $\bar{\eta}$ & $\bar{\eta} = 0.6$ & Midpoint of range of possible values. The setup without imperfections uses this as the fixed efficiency for both detectors.\\
    \hline
    Maximum dark count rate $\dmax$ & $\dmax = 10^{-8}$ & The setup without imperfections assumes the dark count rate to be $0$.\\
    \hline
    Probability of Z-basis preparation & $p_Z=0.9$ & \NA \\
    \hline
    Decoy intensity & $0.01$ & Only one decoy intensity (total two intensities) used.\\
    \hline
    Signal intensity & \NA & Numerical optimisation to find value.\\
    \hline
    Probability of choosing signal intensity & $1/2$ & \NA\\
    \hline
    \end{tabular}
    \caption{Simulation parameters used for secret key rate plots. We do not optimise all the values.}
    \label{tab:sim-params}
\end{table}

\begin{figure*}[ht]
    \centering
    \includegraphics[width = \linewidth]{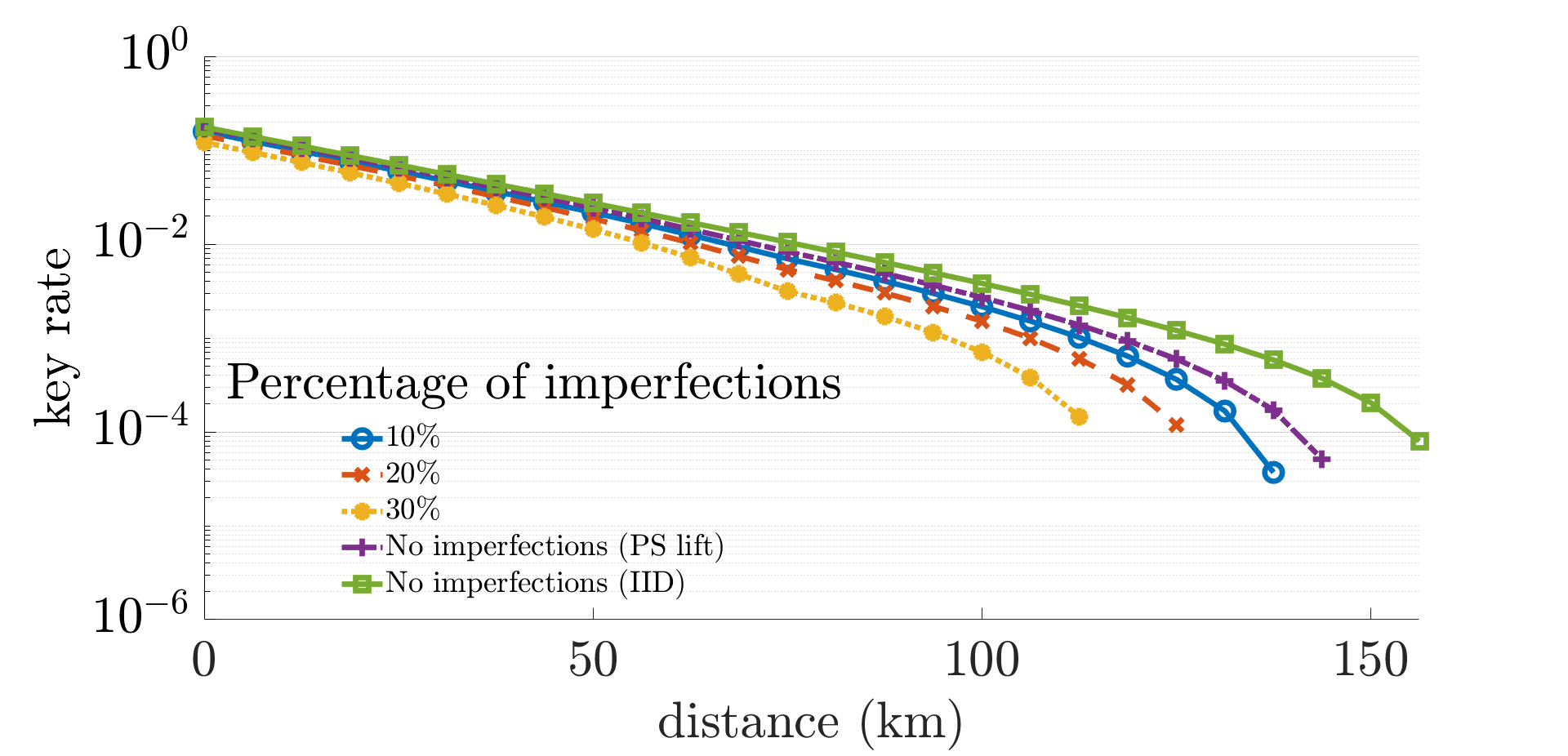}
    \caption{Plots demonstrating the effect of imperfections on the key rate. Our security analysis is minimally affected by detector imperfections. We observe non-zero key rates for imperfection values upto 50\%, though we only plot the key rates upto imperfection values upto 30\%.} \label{fig:impPlot}
\end{figure*}

We consider the beam-splitting ratio and efficiencies to be in varying intervals about the fixed midpoints, $\bar{t} = 0.2$ and $\bar{\eta} = 0.6$, respectively.
We simultaneously vary the ranges of the beam-splitting ratio and detector efficiencies, i.e., we consider the beam-splitting ratio and detector efficiencies to be in ranges 
\begin{equation}
    \begin{gathered}
        t\in[\bar{t}(1-\Delta),\bar{t}(1+\Delta)]\\
        \eta_i \in [\bar{\eta}(1-\Delta),\bar{\eta}(1+\Delta)],
    \end{gathered}
\end{equation}
for the same parameter deviation $\Delta$.
In \cref{fig:impPlot}, we plot\footnote{The code used to generate the plot was produced using the Open QKD
Security package and can be found at https://openqkdsecurity.wordpress.com/repositories-for-publications/.} the key rates for parameter deviations up to $\Delta=30\%$.
We set the maximum dark count rate to be $10^{-8}$.

These results demonstrate that our noise channel framework can tolerate significant detector imperfections while maintaining near-optimal key rates. Part of this high performance is due to the noise channel construction of \cref{thm:genericNoiseChannel}, where the penalty for imperfections is incurred only on the detected rounds. Thus, our treatment of detector imperfections scales favourably with loss. However, we do not fully understand why the security proof is so highly resistant to imperfections. This could be a feature of this specific protocol. A deeper understanding of the techniques and more examples are needed before we can make any conclusive statements.

\subsection{EAT for detectors with memory} \label{sec:corrDetEAT}

We begin our discussion of memory effects in detectors by considering imperfections such as afterpulsing and detector dead times. These effects cause the measurement in one round to depend on detection events in previous rounds \cite{migdall2013single,cusini2022historical}. We assume a finite correlation length $l_c$. This assumption can be justified by using the methodology of \cite[Section 3.3]{narasimhachar2011study} to reduce the problem of infinite-range correlations to one of a finite, effective correlation length with a cost to the security parameter (see \cite[Chapter 7]{nahar_phd_2025} for a more detailed discussion, as well as a different way to address correlated detection setups with noise channels and \cite{wang2025phase} for an alternate way to address correlated detection setups within phase error estimation-based proof techniques).
Specifically, we assume that if a detection event occurs in round \( i \), then the POVM in rounds \( i+1, \dots, i+l_c \) may be altered as a result. In contrast, we assume that if there is no detection in round \( i \), the POVM in rounds \( i+1, \dots, i+l_c \) remain unaffected and behaves as if the measurements were independent. In the following discussion, for pedagogical clarity we set the correlation length to one —-- that is, only the round immediately following a detection event is affected. The extension to larger finite correlation lengths is straightforward.

The main idea is to modify the protocol by discarding any round that is immediately preceded by a detection event. That is, if a detection occurs in round $i$, round $i+1$ is discarded; if round $i+1$ also contains a detection, round $i+2$ is discarded as well, and so on. Intuitively, this eliminates rounds that may exhibit biased or correlated click patterns due to detector memory effects. After this discarding step, all the remaining rounds are the ones that have been measured using POVMs unaffected by correlations. We will outline a proof sketch for such a modified protocol, deferring the full technical treatment to future work.

The main non-trivial task in QKD is to lower bound an entropic quantity\footnote{There are a number of subtleties we have glossed over here for the sake of brevity, such as conditioning on the protocol accepting. Additionally, there exist alternative proof techniques involving different entropic quantities that could be used. For a broader discussion of the steps involved in constructing a full security proof, as well as the different choices available, we refer the interested reader to \cite{tupkary2025qkd} for a more complete description of the various steps that must be taken into account when constructing a full security proof, and the different choices that can be made to do so. For a more complete description specific to the MEAT, we direct the interested reader to \cite{arqand2025marginal}.} $H_{\alpha}^\uparrow(S_1^n\vert C_1^n E_n)$, which can be used with the R\'enyi leftover hashing lemma \cite{dupuis2023privacy} to obtain the secure key length for the QKD protocol. Here, we use the notation $X_1^n$ to denote the $n$ registers $X_1 X_2 \dots X_n$, $C_1^n$ stores the classical announcements made during the protocol, $S_1^n$ stores the key bits, and $E_n$ is the register storing Eve's side-information at the end of the protocol.

Our treatment is specific to the EAT, specifically the marginal-constrained entropy accumulation theorem (MEAT) \cite{arqand2025marginal}, which reduces the task of lower bounding the $n$-round entropic quantity $H_{\alpha}^\uparrow(S_1^n\vert C_1^n E_n)$ to that of lower bounding a single-round quantity that, as semantically implied, can be described in terms of a single protocol round.
Thus, we first informally describe the abstract statement of the MEAT, before sketching the application of the MEAT to IID QKD protocols, i.e. protocols with IID detection setups (with no restrictions on Eve's attack). Finally, we introduce the modifications necessary to handle detector correlations.

\subsubsection{Abstract MEAT statement} \label{subsec:AbstractMEAT}

We make use of \cite[Theorem 4.1a.]{arqand2025marginal} and the proof of \cite[Theorem 4.2a.]{arqand2025marginal}, which considers a sequence of channels depicted in \cref{fig:EATChannels} (we model a QKD protocol in terms of these channels in \cref{subsec:MEATIIDQKD}). The registers \( S_j \) represent secret information, while \( E_j \) and \( C_j \) correspond to a quantum adversarial register and publicly announced classical registers, respectively.

\begin{figure*}
    \centering
    \scalebox{0.8}{\begin{tikzpicture}[node distance=1.5cm and 1.5cm, auto]

\node (box1) [draw, minimum width=1.5cm, minimum height=2cm, fill=gray!20] {\(\mathcal{M}_1\)};
\node (box2) [draw, minimum width=1.5cm, minimum height=2cm, right=2.2cm of box1, fill=gray!20] {\(\mathcal{M}_2\)};
\node (dots) [right=1.3cm of box2] {\(\cdots\)};
\node (boxn) [draw, minimum width=1.5cm, minimum height=2cm, right=1.6cm of dots, fill=gray!20] {\(\mathcal{M}_n\)};

\node (E0) [left=1.2cm of box1] {\({A'}_1^n\)};
\draw[->] (E0) -- (box1.west);

\node (S1) [below=1.2cm of box1] {\(S_1\)};
\node (S2) [below=1.2cm of box2] {\(S_2\)};
\node (Sn) [below=1.2cm of boxn] {\(S_n\)};
\draw[->] (box1.south) -- (S1);
\draw[->] (box2.south) -- (S2);
\draw[->] (boxn.south) -- (Sn);

\node (A1) [above=1.2cm of box1] {$A_1$};
\node (A2) [above=1.2cm of box2] {$A_2$};
\node (An) [above=1.2cm of boxn] {$A_n$};
\draw[->] (A1) -- (box1.north);
\draw[->] (A2) -- (box2.north);
\draw[->] (An) -- (boxn.north);

\draw[->] ([yshift=0.4cm]box1.east) -- node[above] {\(E_1\)} ([yshift=0.4cm]box2.west);
\draw[->] ([yshift=-0.4cm]box1.east) -- node[below] {\(C_1\)} ([yshift=-0.4cm]box2.west);

\draw[->] ([yshift=0.4cm]box2.east) -- node[above] {\(E_2\)} ([yshift=0.4cm]dots.west);
\draw[->] ([yshift=-0.4cm]box2.east) -- node[below] {\(C_2\)} ([yshift=-0.4cm]dots.west);

\draw[->] ([yshift=0.4cm]dots.east) -- node[above] {\(E_{n-1}\)} ([yshift=0.4cm]boxn.west);
\draw[->] ([yshift=-0.4cm]dots.east) -- node[below] {\(C_{n-1}\)} ([yshift=-0.4cm]boxn.west);

\draw[->] ([yshift=0.4cm]boxn.east) -- ++(1.0cm, 0) node[right] {\(E_n\)};
\draw[->] ([yshift=-0.4cm]boxn.east) -- ++(1.0cm, 0) node[right] {\(C_n\)};

\end{tikzpicture}}
    \caption{Sequence of channels used in the abstract MEAT statement in \cref{subsec:AbstractMEAT}.} \label{fig:EATChannels}
\end{figure*}

This can be better understood by stating the final goal of the MEAT theorem --- to lower bound the conditional R\'enyi entropy of the output of the sequence of channels depicted in \cref{fig:EATChannels}, $H_{\alpha}^\uparrow(S_1^n\vert C_1^n E_n) \gtrapprox \sum_i h_{\vert c_1^{j-1}}(c_j)$ in terms of single-round quantities $h_{\vert c_1^{j-1}}(c_j)$ which only depend on a single channel $\mathcal{M}_j$. (Note that $\sum_i h_{\vert c_1^{j-1}}(c_j)$ in the approximate inequality above is more accurately thought of as a form of statistical estimator for the conditional R\'enyi entropy rather than a lower bound. This can be used in the security proof as in \cite{kamin2025r,tupkary2026rigorouscompletesecurityproof}.) Here, each $c_j$ is a particular public announcement stored in the register $C_j$.
While the precise definition of this entropy is not essential for our purposes, the placement of the registers plays an important role in its interpretation. The entropic quantity bounds the amount of information the ``public'' registers $E_n$ and $C_1^n$ have about the ``secret'' registers $S_1^n$ at the end of the sequence of channels.
Importantly, the single-round quantities $h_{\vert c_1^{j-1}}(c_j)$ can be different for different rounds. Further, the functions can also depend on past public announcements $c_1^{j-1}$.

To summarise, the key features of the abstract theorems are as follows:\newline
\textbf{Main result of MEAT:} $H_{\alpha}^\uparrow(S_1^n\vert C_1^n E_n) \gtrapprox \sum_i h_{\vert c_1^{j-1}}(c_j)$
\begin{itemize}
    \item The multi-round entropic term needs to be lower bounded. Here, $S_1^n$ can be understood to be secret registers, while $C_1^n$ and $E_n$ can be understood to be public/adversarial registers.
    \item The single-round quantity can be different for different rounds.
    \item The single-round quantity can depend on past announcements $c_1^{j-1}$.
\end{itemize}

\subsubsection{MEAT for IID protocols} \label{subsec:MEATIIDQKD}

We can now describe the application of \cite[Theorem 4.1a.]{arqand2025marginal} (and the proof of \cite[Theorem 4.2a.]{arqand2025marginal}) to IID QKD protocols, also detailed in \cite[Section 5]{arqand2025marginal}. We model prepare-and-measure QKD protocols via the source-replacement scheme \cite{curty2004entanglement,ferenczi2012symmetries}, which allows us to equivalently think of Alice's state preparation as Alice measuring her part $A_j$ of an entangled state. Thus, the initial state input into the first channel $\mathcal{M}_1$ is an entangled state $\rho_{A_1^n {A'}_1^n}$ given by the source replacement scheme.

Each channel $\mathcal{M}_j$ in round $j$ does the following:
\begin{enumerate}
    \item Measure Alice's system $A_j$ which represents state preparation.
    \item Perform Eve's attack on the system $E_{j-1}$ before sending a register $B_j$ to Bob. Note that this attack in the first round acts on Alice's signal states ${A'}_1^n$ and has no restrictions --- Eve can attack the registers in any order, correlated in any manner.
    \item Measure Bob's system $B_j$ with his POVM $\infPOVM$.
    \item Announcements (such as detect/no detect) made in this round of the protocol are stored in register $C_j$.
    \item Process Alice's measurement results together with the announcements to obtain the secret register $S_j$, which corresponds to key bits.
    \item Updates Eve's side-information register $E_j$ as a result of her attack and the announcements made and passes it to the next channel.
\end{enumerate}

With the above model in mind, the task for a QKD security proof is to bound the conditional entropy $H_{\alpha}^\uparrow(S_1^n\vert C_1^n E_n)$, which can be used with the R\'enyi leftover hashing lemma \cite{dupuis2023privacy} to obtain the maximum possible key length that would be secret from Eve. As described in \cref{subsec:AbstractMEAT}, \cite[Theorem 4.1a.]{arqand2025marginal} allows us to reduce the task of lower bounding the entropy into estimating single-round quantities, which is more tractable \cite{kamin2025r,tupkary2026rigorouscompletesecurityproof} (see \cite{chung2025generalized,matsuura2025asymptotically, he2025operator} for similar bounds on related quantities as well).

\subsubsection{MEAT for detectors with memory}

We finally describe the application of \cite[Theorem 4.1a.]{arqand2025marginal} (and the proof of \cite[Theorem 4.2a.]{arqand2025marginal}) to QKD protocols using detectors with memory. As described in \cref{subsec:MEATIIDQKD}, we once again model prepare-and-measure QKD protocols via the source-replacement scheme \cite{curty2004entanglement,ferenczi2012symmetries}, which allows us to equivalently think of Alice's state preparation as Alice measuring her part $A_j$ of an entangled state. Thus, the initial state input into the first channel $\mathcal{M}_1$ is an entangled state $\rho_{A_1^n {A'}_1^n}$ given by the source replacement scheme.

We describe the action of each channel $\mathcal{M}_j$ in round $j$ as follows (marking the changes from \cref{subsec:MEATIIDQKD} in italics):
\begin{enumerate}
    \item Measure Alice's system $A_j$ which represents state preparation.
    \item Perform Eve's attack on all systems ${A'}_1^n$ before sending a register $B_j$ to Bob. Note that this attack in the first round acts on Alice's signal states ${A'}_1^n$ and has no restrictions --- Eve can attack the registers in any order, correlated in any manner.
    \item Measure Bob's system $B_j$ with his POVM $\infPOVM$ \textit{(this is the POVM that Bob would apply assuming that there are no detector correlations)}.
    \item Announcements (such as detect/no detect) made in this round of the protocol are stored in register $C_j$. \textit{The detect/no detect announcement must be added to the register $C_j$. Note that this announcement need not be made at this stage of the protocol. However, to use the abstract theorem this announcement can always be added to the register $C_j$.}
    \item Process Alice's measurement results together with the announcements to obtain the secret register $S_j$, which corresponds to key bits. \textit{If the previous round detect/no detect announcement stored in $c_{j-1}$ is a detect, then replace the secret register bit with a discard symbol.}
    \item Updates Eve's side-information register $E_j$ as a result of her attack and the announcements made and passes it to the next channel.
\end{enumerate}

This structure fits directly into the abstract MEAT theorem as described in \cref{subsec:AbstractMEAT}, and thus the requisite lower bound on the multi-round entropic quantity can be reduced to a bound on the single-round quantities. We now use the fact that the single-round quantities can depend on past announcements as follows.
The single-round quantity can be set to be $0$ conditioned on the previous round being a detection event (this can be thought of as equivalent to obtaining a no-detection event in the present round). The single-round quantity conditioned on the previous round being a no-detection event can be computed (as described in Ref.~\cite{kamin2025r}) similarly to how it would be if the detectors had no correlations.

We emphasize that the discussion here is only a proof sketch and does not constitute a complete proof. Filling in all technical details—including developing a more physically motivated model of the correlations and performing the necessary calculations—is a highly non-trivial task. We therefore leave this to future work, with the hope that this sketch can serve as a foundation for a full, rigorous proof.

\section{Conclusion}

In this work, we develop a general framework for rigorously addressing imperfections in detection setups. We accomplish this by constructing a channel that `gives' the imperfection to the adversary Eve. The portion `given' to Eve is formalised through a classical `tagged' space as in the flag-state squasher \cite{zhang_security_2021}. We have phrased our results generically to facilitate their application to a variety of adversarial tasks beyond the immediate use in QKD security proofs.
While our theoretical results can be stated generically (see \cref{thm:genericNoiseChannel}), the usage of this theorem requires some notion of an ideal POVM $\finPOVM[\ideal]$, and the estimation of the deviation from ideality $q$ as detailed in \cref{subsubsec:NoiseChannelGenSetups}.
This involves a protocol-dependent computation (see \cref{subsec:EATActiveBB84,subsec:3stateExample} for examples of such a computation).

We further develop protocol-independent theorems for physically motivated models for threshold detectors that include dark counts and non-unit efficiencies.
Specifically, we assume that for all detectors used in the detection setup the dark count rates and efficiencies have specific values (possibly chosen by Eve) within some given range. We emphasise that such a range is typically obtained in characterisation experiments (for e.g. as directed in \cite[Table 4.3]{chunnilall2016quantum}). Thus, our specialised theorems for dark counts (\cref{thm:FSSFineGrained}) and loss (\cref{thm:lossNoiseChannel}) addresses this practically relevant situation. The amount `given' to Eve in this case depends on the maximum dark count rate and the difference in the maximum and minimum efficiency of all the detectors used in the detection setup.

We have also detailed the application of our work to a variety of QKD security proof techniques. Specifically, we describe the application of our results when applied to EAT-based techniques in \cref{subsec:EATActiveBB84}, and we describe the steps required to apply our results to the postselection technique in \cref{sec:appToPS}. In particular, this work can very straightforwardly extend the analyses in Refs.~\cite{kamin2025r,tupkary2026rigorouscompletesecurityproof}. We further, illustrate the impact of our methods on secret key rates for the time-bin encoded three-state protocol in \cref{subsec:3stateExample}. We see that our proof technique can tolerate a very large amount of imperfection.
While we do not explicitly apply our results to proof techniques based on phase error estimation here, this is also possible. Such an application \cite[Chapter 6]{nahar_phd_2025} recovers the asymptotic key rates in \cite[Theorem 7]{gottesman_security_2004}, which improves on past work \cite{tupkary2024phaseerrorrateestimation}. We expect to observe a similar improvement over other work \cite{wang2025phase} on passive detection setups. The impact of our improvements for QKD security proofs is summarised in \cref{tab:imperfect-threshold-setups}.

An important direction for future research is the rigorous treatment of memory effects in detection setups, such as those arising from dead times and afterpulsing. While our main results assume memoryless detectors, we took a first step toward addressing this limitation by outlining a proof strategy in \cref{sec:corrDetEAT} for incorporating detector memory effects into a QKD security proof. Our approach, based on the MEAT developed in Ref.~\cite{arqand2025marginal}, marks the first\footnote{After the initial draft of this work, Ref.~\cite{wang2025phase} developed a full proof for correlated detection setups within phase error estimation-based proof techniques.} explicit treatment of such effects within a formal security analysis. Although we provide only a proof sketch, and the full technical analysis remains non-trivial, we believe that our approach lays a clear foundation for completing such a treatment, and should enable further progress toward a rigorous incorporation of memory effects into QKD protocols.

\section*{Author Contributions}

S.N. developed the main framework for \cref{sec:NoiseChannelWithFSS,sec:appToQKD} and drafted the manuscript. D.T. had the main idea for the protocol change in \cref{sec:corrDetEAT} and contributed to the results in \cref{sec:corrDetEAT}. N.L. formulated the problem, solved the example in \cref{subsec:pedExamp}, and supervised the project. All authors provided feedback on the manuscript.

\begin{acknowledgments}
    We thank Joe Itoi for discussions regarding \cref{subsec:pedExamp}.
    We thank Matej Pivoluska for comments on an earlier draft.
    We thank Lars Kamin, Amir Arqand, and Ernest Tan for discussions regarding \cref{sec:corrDetEAT} and the MEAT.
    We thank Zhiyao Wang for suggesting an improvement for \cref{eq:subspaceWtIncreaseLoss} and the noise channel construction in \cref{subsec:noiseChannelThreeState}.
    The work has been performed at the Institute for
    Quantum Computing, at the University of Waterloo,
    which is supported by Innovation, Science, and Economic
    Development Canada. The research has been supported
    by NSERC under the Discovery Grants Program, Grant
    No. 341495.
    This research has been supported by Alliance QUINT.
    DT is funded by the Mike and Ophelia Lazaridis Fellowship.
\end{acknowledgments}

\appendix
\section{Numerically checking the existence of a noise channel} \label{app:numCheck}

In this appendix we describe a numeric method to check \cref{eq:PPswap,eq:noiseChannelCond} for generic protocols with any squashing map, but for a specific value of the dark count rates $\dB$ and loss $\loss$.
Since it requires a specific value of the dark count rates $\dB$ and loss $\loss$ (and although the analysis can be repeated a finite number of times for different values of $\dB$, $\loss$), this method does not constitute a rigorous treatment of untrusted dark counts. However, it serves as a useful tool when searching for a possible post-processing that satisfies equation \cref{eq:PPswap}, and a possible noise channel satisfying \cref{eq:noiseChannelCond}.

\subsection{\cref{eq:PPswap} as a linear program} \label{app:PPswapLP}

\begin{align*}
    \Msq' \PPdb = \Mdc\Msq
\end{align*}
Under the simplifying assumption that $\Msq' = \Msq$, it is clear that this is just a linear constraint for the variable $\Mdc$. In addition to this constraint, we need to constrain it to be a stochastic matrix, i.e., all entries need to be positive and each row needs to sum to 1. As these are all linear constraints, this is a linear program (where we only need to find a feasible point).

\subsection{\cref{eq:noiseChannelCond} as a semidefinite program}

\begin{align*}
    \Mdc\Tr[\finPOVM[\loss] \rho] = \Tr[\finPOVM\ \noiseChannel(\rho)] \quad \quad \forall\; \rho
\end{align*}
Here $\Mdc$ is obtained from \cref{app:PPswapLP}.
This can be rephrased using the Choi isomorphism $J$ of the noise channel $\Phi$ as
\begin{align} \label{eq:noiseChannelCondChoiMap}
    \Mdc\Tr[\finPOVM[\loss] \rho] = \Tr\left[\left(\rho^\intercal \otimes \finPOVM\right) J\right] \quad \quad \forall\; \rho.
\end{align}
It is sufficient to satisfy \cref{eq:noiseChannelCondChoiMap} on a (finite) set of $\rho$ that constitutes a basis for the space. This gives a finite set of linear constraints on the Choi map $J$.
Additionally, since $J$ is the Choi representation of a channel, it must be postive semidefinite, and tracing over the second space must give the identity on the first space.
All of these are positive semidefinite or linear constraints and thus this constitutes a semidefinite program (where we only need to show there exists a feasible point).

As we are not using this as a rigorous proof technique --- we only use it as an intuition building tool --- we do not worry about numerical imprecision.

\section{Noise channel constructions} \label{app:noiseChannelProofs}

\FSSFineGrained*
\begin{proof}
    The existence of the flag-state squasher is directly proved in \cite[Theorem 1]{zhang_security_2021}. Note that since $\H_0$ is one-dimensional, this can also serve as the flag for the no-click event.
    We make an ansatz for the noise channel, depicted in \cref{fig:noiseChannelFSS} which we will now elaborate on.
    The noise channel first projects the input state into $\H_1$ and $\H_0\oplus\H_F$. On $\H_0\oplus\H_F$, it first measures the state with the lossy POVM $\finPOVM[\loss]$, and then applies the dark count post-processing map on the measurement results.
    
    On $\H_1$, it probabilistically chooses between two channels.
    With probability $\PP{0}{0}^\dB$, it acts as the identity channel. Intuitively, this corresponds to the case when there are no dark counts in the setup. When there are dark counts, we assign the detection events to classical states in the flag-space as follows.
    With probability $1-\PP{0}{0}^\dB$, the noise channel measures the state with $\finPOVM[\loss]$ to obtain result $\vec{p} = \Tr[\finPOVM[\loss] \rho_1]$, where $\rho_1$ is the part of $\rho$ acting on $\H_1$. The noise channel then uses the measurement result $\vec{p}$ to prepare the classical state
    \begin{equation}
        \classState = \frac{1}{1-\PP{0}{0}^\dB}\left(\sum_{\substack{m\in\multiSet\\ j\notin \multiSet}} \PP{m}{j}^\dB p_j \ketbra{m} + \sum_{s\in\singleSet} \left(\PP{s}{0}^\dB p_0 + (\PP{s}{s}^\dB-\PP{0}{0}^\dB)p_s\right)\ketbra{s}\right).
    \end{equation}

    \begin{figure*}[h]
        \centering
        \scalebox{0.75}{\begin{tikzpicture}[scale = 1.25]

\def\mylinewidth{2pt}

\coordinate (h1) at (4,0);
\coordinate (h2) at (12,0);
\coordinate (h3) at (0,-1);
\coordinate (h4) at (2,-1);
\coordinate (h5) at (4,-2);
\coordinate (h6) at (9,-2);
\coordinate (h7) at (10,-3);
\coordinate (h8) at (12,-3);
\coordinate (h9) at (0,-4);
\coordinate (h10) at (9,-4);
\pic [right of = h9,xshift = 4cm,scale = 0.78]{detector={flagMeas, $\finPOVM[\loss]$, black}};
\node[process,right of = flagMeas,xshift = 2cm,scale = 1.5](flagPP){$\PP^\dB$};
\pic [right of = h5,xshift = 0.5cm,scale = 0.75]{detector={meas, $\finPOVM[\loss]$,black}};
\pic[right of = meas, xshift = 0.5cm, scale = 0.5]{source={prep, $\classState$, black,3}};

\node[left of = h3, font=\LARGE](H1){$\H_1$};
\node[left of = h9,xshift = -0.5 cm, font=\LARGE](H0HF){$\H_0\oplus \H_F$};
\node[below of = H1,yshift = -1cm, font = \LARGE]{$\oplus$};
\node[right of = h2, font=\LARGE](H1end){$\H_1$};
\node[right of = h8,xshift = 0.5 cm, font=\LARGE](H0HFend){$\H_0\oplus \H_F$};
\node[below of = H1end,yshift = -1cm, font = \LARGE]{$\oplus$};

\node[fit=(h1)(h4)(h5)(h6)(h7)(h10)(flagPP), draw, line width = \mylinewidth*2, inner sep=15pt](fitrect) {};

\draw[cyan, line width=\mylinewidth] (h1) -- (h2);
\draw[cyan, line width=\mylinewidth] (h3) -- (h4);
\draw[cyan, line width=\mylinewidth] (h5) -- ([xshift=-0.11cm]meas.west);
\draw[dashed, cyan, line width=\mylinewidth] ([xshift=0.16cm]meas.east) -- ([xshift=-0.27cm]prep.west);
\draw[dashed, cyan, line width=\mylinewidth] ([xshift=0.66cm]prep.east) -- ([xshift = -0.08 cm]h6);

\draw[dashed, cyan, line width=\mylinewidth] (h7) -- (h8);
\draw[cyan, line width=\mylinewidth] (h9) -- ([xshift=-0.12cm]flagMeas.west);
\draw[dashed, cyan, line width=\mylinewidth] ([xshift=0.18cm]flagMeas.east) -- (flagPP);

\draw[dashed, cyan, line width=\mylinewidth] (flagPP) -- (h10);

\draw[cyan, line width=\mylinewidth] (h4) -- (h1) node[midway, above, yshift = 0.1cm, black]{$\PP{0}{0}^\dB$};
\draw[cyan, line width=\mylinewidth] (h4) -- (h5) node[midway, below, yshift = -0.3cm, black]{$1-\PP{0}{0}^\dB$};
\draw[dashed, cyan, line width=\mylinewidth] (h6) -- (h7);
\draw[dashed, cyan, line width=\mylinewidth] (h10) -- (h7);

\node[above=0.5cm of fitrect, font=\LARGE] (Title) {\textbf{Noise Channel} $\noiseChannel{\dB}$};

\end{tikzpicture}}
        \caption{Constructive description of the noise channel. Each line corresponds to a subspace of the input Hilbert space associated with the block-diagonal decomposition of the POVM elements. The dashed lines refer to classical states.} \label{fig:noiseChannelFSS}
    \end{figure*}

    The rest of the proof follows by proving the following:
    \begin{enumerate}
        \item $\Phi$ is completely-positive and trace-preserving (CPTP),
        \item $\PPdb\Tr[\finPOVM[\loss]\rho] = \Tr[\finPOVM[\loss]\ \Phi(\rho)]$ for all input density matrices $\rho$,
        \item $\Tr[\Phi(\rho)\Pi_{\leq 1}] = \PP{0}{0}^\dB\Tr[\rho\Pi_{\leq 1}]$ for all $\rho$.
    \end{enumerate}
    \textbf{$\Phi$ is CPTP:}
    Since it is clear that measurements and classical post-processing are physical channels, $\Phi$ can be shown to be CPTP by showing that $\classState$ is a valid classical state as the CPTP property is preserved under channel composition. Note that all the elements of $\classState$ are positive as all elements of the stochastic matrix $\PPdb$ are positive, and $\PPdb{s}{s} \geq \PP{0}{0}$ for all $s\in\singleSet$ as described in \cref{eq:11morethan00}. Also, note that $\sum_{j\notin \multiSet} p_j =1$ as $p_m= 0$ for all $m\in \multiSet$ as stated in \cref{eq:noSingleToMulti}. Then we compute,
    \begin{align}
        \nonumber\Tr[\classState] &= \frac{1}{1-\PP{0}{0}^\dB}\left(\sum_{\substack{m\in\multiSet\\ j\notin \multiSet}} \PP{m}{j}^\dB p_j + \sum_{s\in\singleSet} \left(\PP{s}{0}^\dB p_0 + (\PP{s}{s}^\dB-\PP{0}{0}^\dB)p_s\right)\right)\\
        \nonumber &= \frac{1}{1-\PP{0}{0}^\dB}\left(\sum_{\substack{s\in\singleSet\\m\in\multiSet}}\left(\PP{m}{0}^\dB+\PP{s}{0}^\dB \right)p_0+ \sum_{\substack{s\in\singleSet\\m\in\multiSet}}\left(\PP{m}{s}^\dB+\PP{s}{s}^\dB-\PP{0}{0}^\dB\right)p_s\right)\\
        \label{eq:TPProofStep} &= \frac{1}{1-\PP{0}{0}^\dB}\left(\left(1-\PP{0}{0}^\dB\right)p_0 + \sum_{s\in\singleSet} \left(1-\PP{0}{0}^\dB\right)p_s\right)\\
        \nonumber &= 1,
    \end{align}
    where \cref{eq:TPProofStep} follows from the fact that $\sum_{\substack{m\in\multiSet}}\PP{m}{s}^\dB+\PP{s}{s}^\dB = 1$ for any $s\in\singleSet$ as implied by \cref{eq:sIsS,eq:noWhiteCounts}.

    \textbf{$\PPdb\Tr[\finPOVM[\loss]\rho] = \Tr[\finPOVM[\loss]\ \Phi(\rho)]$ for all input density matrices $\rho$:} Due to the block-diagonal structure of the POVM elements, it suffices to show that this condition holds for flag-states $\{\ketbra{i}\}_{i=1}^\nmeas$, vacuum state $\ketbra{\vac}$ and single-photon states $\rho_1$ separately. Any input state can then be written as a convex combination of these states, and linearity would extend this for all states.
    
    For any flag-state $\ketbra{i}$, the condition $\PPdb\Tr[\finPOVM[\loss]\ketbra{i}] = \Tr[\finPOVM[\loss]\ \Phi(\ketbra{i})]$ follows directly from the construction depicted in \cref{fig:FSSEquivalences}. This similarly holds true for the vacuum state. We now turn our attention to the single-photon states $\rho_1$.
    We divide the computation for the single-photon states into three parts.
    First, consider the statistics of all $m \in \multiSet$. Using \cref{eq:noSingleToMulti}, we have that
    \begin{align}
        \nonumber\Tr[F^{\loss}_m \, \Phi(\rho_1)] &= \PP{0}{0}^\dB \Tr[F^{\loss}_m\rho_1] + (1-\PP{0}{0}^\dB)\Tr[F^{\loss}_m\classState]\\
        \label{eq:multiSetNoiseCondStep1}&= 0 + \sum_{j\notin \multiSet} \PP{m}{j}^\dB \Tr[F^{\loss}_j \rho_1]\\
        \label{eq:multiSetNoiseCondStep2}&= 
        \sum_{j\in U} \PP{m}{j} \Tr[F^{\loss}_j \rho_1],
    \end{align}
    where \cref{eq:multiSetNoiseCondStep1,eq:multiSetNoiseCondStep2} follow from \cref{eq:noSingleToMulti}.
    
    Next, we consider the statistics of all $s\in\singleSet$.
    \begin{align}
        \nonumber\Tr[F^{\loss}_s \, \Phi(\rho_1)] &= \PP{0}{0}^\dB \Tr[F^{\loss}_s\rho_1] + (1-\PP{0}{0}^\dB)\Tr[F^{\loss}_s\classState]\\
        \nonumber &= \PP{0}{0}^\dB \Tr[F^{\loss}_s\rho_1] + \left(\PP{s}{0}^\dB \Tr[F^{\loss}_0\rho_1] + (\PP{s}{s}^\dB-\PP{0}{0}^\dB)\Tr[F^{\loss}_s \rho_1]\right)\\
        \nonumber &= \PP{s}{0}^\dB \Tr[F^{\loss}_0\rho_1] + \PP{s}{s}^\dB\Tr[F^{\loss}_s \rho_1]\\
        \label{eq:singeSetNoiseCondStep} &= \sum_{j\in U} \PP{s}{j}^\dB \Tr[F^{\loss}_j \rho_1],
    \end{align}
    where \cref{eq:singeSetNoiseCondStep} follows from \cref{eq:sIsS,eq:noSingleToMulti}.
    
    Finally, we consider the no-click statistics,
    \begin{align}
        \nonumber \Tr[F^{\loss}_0 \; \Phi(\rho_1)] &= \PP{0}{0}^\dB \Tr[F^{\loss}_0\rho_1] + (1-\PP{0}{0}^\dB)\Tr[F^{\loss}_0\classState]\\
        &= \PP{0}{0}^\dB \Tr[F^{\loss}_0\rho_1]+0\\
        &= \sum_{j\in U} \PP{0}{j}^\dB \Tr[F^{\loss}_j \rho_1],
    \end{align}
    where the last equality follows from \cref{eq:noWhiteCounts}.
    Thus, the noise channel $\Phi$ is a CPTP map that satisfies \cref{eq:noiseChannelCond}.

    \textbf{$\Tr[\Phi(\rho)\Pi_{\leq 1}] = \PP{0}{0}^\dB\Tr[\rho\Pi_{\leq 1}]$ for all $\rho$:} Given some density matrix $\rho$, we let $\Pi_{\leq 1} \rho \Pi_{\leq 1} = q_0 \ketbra{\vac} + q_1 \rho_1$. From the construction depicted in \cref{fig:noiseChannelFSS}, we see that $\Phi(\ketbra{\vac}) = \PP{0}{0}^\dB\ketbra{\vac} + \sum_{i\neq 0} \PP{i}{0}^\dB\ketbra{i}$. Moreover, $\Phi(\rho_1) = \PP{0}{0}^\dB \rho_1 +\classState$, where $\classState$ is composed entirely of flags. Finally, $\Phi(\ketbra{i})$ only has support on the flag space $\H_F$. Thus, $\Pi_{\leq 1} \Phi(\rho) \Pi_{\leq 1} = q_0 \PP{0}{0}^\dB \ketbra{\vac} + q_1 \PP{0}{0}^\dB \rho_1$. Since $q_0 + q_1 = \Tr[\Pi_{\leq 1}\rho]$, we have that $\Tr[\Pi_{\leq 1}\Phi(\rho)] = \PP{0}{0}^\dB \Tr[\Pi_{\leq 1} \rho]$.
\end{proof}

\lossNoiseChannel*
\begin{proof}
    The structure of the proof is very similar to the proof of \cref{thm:FSSFineGrained}.
    First, the existence of the flag-state squasher is directly proved in \cite[Theorem 1]{zhang_security_2021}. Let the (lossy) POVM after the application of the flag-state squasher be given by $\finPOVM[\loss]$. Note that this is related to the lossless POVM $\finPOVM$ as
    \begin{align} \label{eq:lossLeq1}
        \finPOVM[\loss, m\leq1] = \PPloss \finPOVM[m\leq1],
    \end{align}
    where
    \begin{align*}
        \PPloss &= \begin{pNiceMatrix}
            1 & 1-\eta_1 & \Cdots &   & 1-\eta_k\\
            0 & \eta_1 & \Cdots &   & 0\\
            \Vdots & \Vdots & \Ddots & &  \Vdots\\
             &  & & \Ddots &   \\
            0 & 0 & \Cdots & &  \eta_k
        \end{pNiceMatrix},
    \end{align*}
    and $\finPOVM[\loss, m\leq1] = \finPOVM[m=0]\oplus\finPOVM[\loss, m=1]$.
    Additionally, the flag-space portion, and the vacuum ($m=0$) portion of both the noisy, and noiseless POVMs are identical.

    The construction of the noise channel for the lossy POVM $\finPOVM[\loss]$ is similar to the construction used in \cref{thm:genericNoiseChannel}.
    For any value of common efficiency $\idealEff \in \left[\frac{\etamin}{1-(\etamax-\etamin)},1\right]$, we break up the loss post-processing $\PPloss$ as
    \begin{align} \label{eq:breakUpLossPP}
        \PPloss = \frac{\etamin}{\idealEff}\PP^{\idealEff} + \left(1-\frac{\etamin}{\idealEff}\right) \mathcal{Q}^{\loss,\idealEff},
    \end{align}
    where
    \begin{align*}
        \mathcal{Q}^{\loss,\idealEff} &= \begin{pNiceMatrix}
            1 & 1-\idealEff\frac{\eta_1-\etamin}{\idealEff-\etamin} & \Cdots &   & 1-\idealEff\frac{\eta_k-\etamin}{\idealEff-\etamin}\\
            0 & \idealEff\frac{\eta_1-\etamin}{\idealEff-\etamin} & \Cdots &   & 0\\
            \Vdots & \Vdots & \Ddots & &  \Vdots\\
             &  & & \Ddots &   \\
            0 & 0 & \Cdots & &  \idealEff\frac{\eta_k-\etamin}{\idealEff-\etamin}
        \end{pNiceMatrix},
    \end{align*}
    Thus, from \cref{eq:lossLeq1,eq:breakUpLossPP} we get
    \begin{align}
        \finPOVM[\loss, m\leq1] = \frac{\etamin}{\idealEff}\PP^{\idealEff} \finPOVM[m\leq1] + \left(1-\frac{\etamin}{\idealEff}\right) \mathcal{Q}^{\loss,\idealEff} \finPOVM[m\leq1].
    \end{align}
    Further, note that $\finPOVM[m=0]$ is independent of the loss $\loss$, so that we can refine the above equation to
    \begin{align}
        \finPOVM[\loss, m\leq1] = \finPOVM[m=0]\oplus\left(\frac{\etamin}{\idealEff}\PP^{\idealEff} \finPOVM[m=1] + \left(1-\frac{\etamin}{\idealEff}\right) \mathcal{Q}^{\loss,\idealEff} \finPOVM[m=1]\right).
    \end{align}
    
    Thus, we can construct the noise channel depicted in \cref{fig:noiseChannelFSSLoss}:
    \begin{align}
        \noiseChannel{\loss}^{\idealEff}\left(\begin{pNiceArray}{c|c|c}
                {\rho_{0}} & {B} & {C}\\
                \hline
                {B^\dag} & {\rho_{1}} & {D}\\
                \hline
                {C^\dag} & {D^\dag} & \rho_{F}
            \end{pNiceArray}\right) \coloneqq 
            \rho_0\oplus\frac{\etamin}{\idealEff}\rho_1\oplus \left(\left(1-\frac{\etamin}{\idealEff}\right) \sum_i \Tr[\rho_1 Q_i] \ketbra{i} + \rho_F\right),
    \end{align}
    where $\vec{Q} \coloneqq \mathcal{Q}^{\loss,\idealEff} \finPOVM[m=1]$.
    It is easy to verify that this construction satisfies $\Tr[\infPOVM[\loss]\rho] = \Tr[\finPOVM[\idealEff]\ \Phi_{\loss}^{\idealEff}\left(\Lambda(\rho)\right)]$, and that \cref{eq:subspaceWtIncreaseLoss} holds.
    \begin{figure*}[h]
        \centering
        \scalebox{0.8}{\begin{tikzpicture}[scale = 1.25]

\def\mylinewidth{2pt}

\coordinate (h0L) at (0,1);
\coordinate (h0R) at (12,1);
\coordinate (h0M) at (6,1);
\coordinate (h1) at (4,0);
\coordinate (h2) at (12,0);
\coordinate (h3) at (0,-1);
\coordinate (h4) at (2,-1);
\coordinate (h5) at (4,-2);
\coordinate (h6) at (9,-2);
\coordinate (h7) at (10,-3);
\coordinate (h8) at (12,-3);
\coordinate (h9) at (0,-4);
\coordinate (h10) at (9,-4);
\pic [right of = h9,xshift = 4cm,scale = 0.78]{detector={flagMeas, $\finPOVM[\loss]$, black}};
\pic [right of = h5,xshift = 0.5cm,scale = 0.75]{detector={meas, $\finPOVM$,black}};
\node[process,right of = meas, xshift = 1.5cm, scale = 1.5](lossPP){$\mathcal{Q}^{\loss,\idealEff}$};

\node[left of = h0L, xshift=0cm, font=\LARGE] (H0L) {$\H_{0}$};
\node[right of = h0R, xshift=0cm, font=\LARGE] (H0R) {$\H_{0}$};
\node[below of = H0L,yshift = -0.25cm, font = \LARGE]{$\oplus$};
\node[below of = H0R,yshift = 0.4cm, font = \LARGE]{$\oplus$};
\node[left of = h3, font=\LARGE](H1){$\H_{1}$};
\node[left of = h9,xshift = 0 cm, font=\LARGE](H0HF){$\H_{m>1}$};
\node[below of = H1,yshift = -1cm, font = \LARGE]{$\oplus$};
\node[right of = h2, font=\LARGE](H1end){$\H_{1}$};
\node[right of = h8,xshift = 0 cm, font=\LARGE](H0HFend){$\H_F$};
\node[below of = H1end,yshift = -1cm, font = \LARGE]{$\oplus$};

\node[fit=(h1)(h4)(h5)(h6)(h7)(h10)(h0M), draw, line width = \mylinewidth*2, inner sep=30pt](fitrect) {};

\draw[cyan, line width=\mylinewidth] (h1) -- (h2);
\draw[cyan, line width=\mylinewidth] (h3) -- (h4);
\draw[cyan, line width=\mylinewidth] (h5) -- ([xshift=-0.11cm]meas.west);
\draw[dashed, cyan, line width=\mylinewidth] ([xshift=0.16cm]meas.east) -- (lossPP.west);
\draw[dashed, cyan, line width=\mylinewidth] (lossPP.east) -- ([xshift = -0.08 cm]h6);

\draw[dashed, cyan, line width=\mylinewidth] (h7) -- (h8);
\draw[cyan, line width=\mylinewidth] (h9) -- ([xshift=-0.12cm]flagMeas.west);
\draw[dashed, cyan, line width=\mylinewidth] ([xshift=0.18cm]flagMeas.east) -- (h10);

\draw[cyan, line width=\mylinewidth] (h0L) -- (h0R);

\draw[cyan, line width=\mylinewidth] (h4) -- (h1) node[midway, above, yshift = 0.2cm, black]{$\frac{\etamin}{\idealEff}$};
\draw[cyan, line width=\mylinewidth] (h4) -- (h5) node[midway, below, yshift = -0.3cm, black]{$1-\frac{\etamin}{\idealEff}$};
\draw[dashed, cyan, line width=\mylinewidth] (h6) -- (h7);
\draw[dashed, cyan, line width=\mylinewidth] (h10) -- (h7);

\node[above=0.5cm of fitrect, font=\LARGE] (Title) {\textbf{Noise Channel $\Phi_\loss^{\idealEff}$}};

\end{tikzpicture}}
        \caption{Constructive description of the noise channel. Each line corresponds to a subspace of the input Hilbert space associated with the block-diagonal decomposition of the POVM elements. The dashed lines refer to classical states.} \label{fig:noiseChannelFSSLoss}
    \end{figure*}
\end{proof}

\section{Bounds on flag space weight for active basis-choice BB84 detection setup} \label{app:flagWtActiveBB84}

In this section, we bound the quantities needed to upper bound the weight in the flag space needed for the security proof of the active basis-choice BB84 in \cref{subsec:EATActiveBB84}. First, we list the POVM used in the protocol.
Specifically, we upper bound
\begin{itemize}
    \item the deviations from ideality $q_0$ and $\qbasis$ which represents the probability of the noise channel to send a round to the flag space, and
    \item the weights in the $0$ and single-photon subspaces in terms of observable quantities.
\end{itemize}

These bounds depend on the POVM elements in each basis:
\begin{equation} \label{eq:detPOVMdef}
    \begin{aligned}
        \GammancBasis^{\dB,\loss} &= \sum_{N_0, N_1=0}^\infty (1-d_0)(1-d_1)(1-\eta_{0})^{N_0} (1-\eta_{1})^{N_1} \ketbra{N_0,N_1}_b\\
        \GammazeroBasis^{\dB,\loss} &= \sum_{N_0, N_1=0}^\infty \frac{1}{2}(1-(1-d_0)(1-\eta_{0})^{N_0})(1+(1-d_1)(1-\eta_{1})^{N_1}) \ketbra{N_0,N_1}_b\\
        \GammaoneBasis^{\dB,\loss} &= \sum_{N_0, N_1=0}^\infty \frac{1}{2}(1+(1-d_0)(1-\eta_{0})^{N_0})(1-(1-d_1)(1-\eta_{1})^{N_1}) \ketbra{N_0,N_1}_b,
    \end{aligned}
\end{equation}
where $d_0$, $d_1$ and $\eta_0$, $\eta_1$ are the dark count rates and efficiencies of the two detectors, and $\ketbra{N_0, N_1}_b$ is the state with $N_0$ photons in the mode 1, and $N_1$ photons in mode 2, where the modes are defined with respect to basis $b$. Here, we have included the classical postprocessing which randomly maps double clicks to the $0$ and $1$ outcome in the same basis into the POVM descriptions.

These POVM elements can be used to construct the full POVM describing the detection setup by multiplying each of them with the probability of picking basis $b$. As this probability is the same for the real and ideal setups, it does not appear in our computations. Thus, for ease of notation we will henceforth refer to the POVMs described in \cref{eq:detPOVMdef}, without the probability prefactor.

\subsection{Bounding deviations from ideality} \label{appsec:noiseChannelDeviationPhaseError}

We now proceed by upper bounding the deviation from ideality $\qbasis$. This is defined such that $\Gamma_{i,\, b,\ >0}^{\dB,\loss} \geq (1-\qbasis)\Gamma_{i,\, b,\ >0}$ for all outcomes $i$ and bases $b$. As both POVM elements corresponding to a specific outcome are diagonal in the same photon-number basis, the operator inequality reduces to a set of scalar inequalities on their corresponding eigenvalues.
Thus, it is sufficient to lower bound the ratio of their eigenvalues\footnote{Note that these eigenvalues do not depend on the basis chosen.} $\lambda^{\dB,\loss}_{i;\;N_0,N_1}$ and $\lambda_{i;\;N_0,N_1}$:
\begin{align}
    \qbasis &\geq 1-\min_{\substack{i\\ N_0+N_1\neq 0}}\left\{ \frac{\lambda^{\dB,\loss}_{i;\;N_0,N_1}}{\lambda_{i;\;N_0,N_1}}\right\}.
\end{align}
The ideal $\Gamma_{>0;\ \nc,\; b}$ is the $0$ operator outside the zero photon subspace, and so the minimisation can be restricted to the $0$ and $1$ outcomes.

We explicitly compute the ratio for outcome $0$. First, note from \cref{eq:detPOVMdef} that 
\begin{equation}
    \begin{aligned}
        \lambda_{0;\;0,N_1} &= 0 \qquad \qquad \forall\; N_1 \neq 0\\
        \lambda_{0;\;N_0,0} & = 1 \qquad \qquad \forall\; N_0 \neq 0\\
        \lambda_{0;\;N_0,0} & = 1/2 \quad \qquad \forall\; N_0, N_1 \neq 0.
    \end{aligned}
\end{equation}$\lambda_{0;\;0,N_1} = 0$ for all $N_1$, $\lambda_{0;\;N_0,0} = 1$ for all $N_0\neq 0$, and $\lambda_{0;\;N_0,0} = 1/2$ for all $N_0, N_1\neq 0$. When $N_0=0$, the ratio diverges. Thus, we compute the ratio for the remaining two cases separately, and then we minimise over all cases after:
\begin{gather}
    \begin{aligned}
        \min_{N_0\neq0}\left\{ \frac{\lambda^{\dB,\loss}_{0;\;N_0,0}}{\lambda_{0;\;N_0,0}}\right\} &= \min_{N_0\neq0}\left\{(1-(1-d_0)(1-\eta_{0})^{N_0})\left(1-\frac{d_1}{2}\right)\right\} \\
        &= (1-(1-d_0)(1-\eta_{0}))\left(1-\frac{d_1}{2}\right)\\
        &\geq \etamin \left(1-\frac{\dmax}{2}\right)
    \end{aligned}\\
    \begin{aligned}
        \min_{N_0,N_1\neq0}\left\{ \frac{\lambda^{\dB,\loss}_{0;\;N_0,N_1}}{\lambda_{0;\;N_0,N_1}}\right\} &= \min_{N_0,N_1\neq0}\left\{(1-(1-d_0)(1-\eta_{0})^{N_0})(1+(1-d_1)(1-\eta_{1})^{N_1})\right\} \\
        &= (1-(1-d_0)(1-\eta_{0}))\\
        &\geq \etamin.
    \end{aligned}
\end{gather}
By a similar computation for outcome $1$, we achieve the same minimum ratio, $\etamin \left(1-\frac{\dmax}{2}\right)$, so that 
\begin{equation*}
    \qbasis = 1- \etamin \left(1-\frac{\dmax}{2}\right).
\end{equation*}

We now upper bound the deviation from ideality $q_0$ corresponding to the $0$-photon subspace. Similar to the previous case, it is sufficient to lower bound the ratio of their eigenvalues
\begin{align}
    q_0 &\geq 1-\min_{\substack{i}}\left\{ \frac{\lambda^{\dB,\loss}_{i;\;0,0}}{\lambda_{i;\;0,0}}\right\}.
\end{align}
The ideal $\Gamma_{\nc,\ 0}$ has eigenvalue 1 in the $0$-photon subspace, and all other ideal POVM elements have eigenvalue 0. Thus, the minimisation reduces to
\begin{align}
    \nonumber q_0 &\geq 1- \frac{\lambda^{\dB,\loss}_{\nc;\;0,0}}{\lambda_{\nc;\;0,0}}\\
    &= 1-(1-d_0)(1-d_1) \geq 1-(1-\dmax)^2
\end{align}

\newcommand{\lambdaminBasis}{\lambda_{\mathrm{min;\,\det}}}

\subsection{Bounding weight outside $0$-photon subspace} \label{appsec:minEigenvaluePhaseError}

We bound the weight outside the $0$-photon subspace by obtaining a lower bound on the minimum eigenvalue $\lambdaminBasis$ of $\Pi_{>0}\Gamma_{\det}\Pi_{>0}^{\dB,\loss} \coloneqq \Pi_{>0} - \Pi_{>0}\Gamma_{\nc}^{\dB,\loss}\Pi_{>0}$, which directly implies $\Gamma_{\det}^{\dB,\loss} \geq \lambdaminBasis\Pi_{>0}$ as needed.
This minimum eigenvalue can be obtained from \cref{eq:detPOVMdef} by directly minimising over all eigenvalues of $\Gamma_{\det}^{\dB,\loss}$ outside the zero-photon subspace,
\begin{equation} \label{eq:lambdaMinActiveEUR}
    \lambdaminBasis = \min_{N_0+N_1 \neq 0}\left\{1-(1-d_0)(1-d_1)(1-\eta_{0})^{N_0} (1-\eta_{1})^{N_1}\right\} \geq \etamin.
\end{equation}

\bibliography{references}
\bibliographystyle{quantum}

\end{document}